\documentclass[12pt,preprint]{emulateapj}

\slugcomment{}

\shorttitle{Photometric redshifts for AGN}
\shortauthors{Salvato et~al.}

\begin{document}


\title{Photometric redshift and classification for the {\it XMM}-COSMOS sources\altaffilmark{*}}

\author{M. Salvato\altaffilmark{1},
G. Hasinger\altaffilmark{2,3},
O. Ilbert\altaffilmark{3},
G. Zamorani\altaffilmark{4},
M. Brusa\altaffilmark{2}, 
N.  Z. Scoville\altaffilmark{1},
A. Rau\altaffilmark{1},
P. Capak\altaffilmark{1,19},
S. Arnouts\altaffilmark{5},
H. Aussel\altaffilmark{6},
M. Bolzonella\altaffilmark{4},
A. Buongiorno\altaffilmark{2},
N. Cappelluti\altaffilmark{2},
K. Caputi\altaffilmark{9},
F. Civano\altaffilmark{21},
R. Cook\altaffilmark{7,1,15},
M. Elvis\altaffilmark{21},
R. Gilli\altaffilmark{18},
K. Jahnke\altaffilmark{20},
J.S. Kartaltepe\altaffilmark{3},
C.D. Impey\altaffilmark{8},
F. Lamareille\altaffilmark{2},
E. Le Floch'\altaffilmark{3},
S. Lilly\altaffilmark{9},
V. Mainieri\altaffilmark{10},
P. McCarthy\altaffilmark{22}.
H. McCracken\altaffilmark{11},
M. Mignoli\altaffilmark{4},
B. Mobasher\altaffilmark{12},
T. Murayama\altaffilmark{24},
S. Sasaki\altaffilmark{13},
D.B. Sanders\altaffilmark{3},
D. Schiminovich\altaffilmark{18},
Y. Shioya\altaffilmark{17},
P. Shopbell\altaffilmark{1},
J. Silvermann\altaffilmark{9},
V. Smol\v{c}i\'{c}\altaffilmark{1},
J. Surace\altaffilmark{19},
Y. Taniguchi\altaffilmark{17},
D. Thompson\altaffilmark{14},
J.R. Trump\altaffilmark{8},
M. Urry\altaffilmark{23},
M. Zamojski\altaffilmark{18,19}
  }
  \altaffiltext{$\star$}{Based on observations with the NASA/ESA {\em
Hubble Space Telescope}, obtained at the Space Telescope Science
Institute, which is operated by AURA Inc, under NASA contract NAS
5-26555. Also based on observations made with the Spitzer Space Telescope,
which is operated by the Jet Propulsion Laboratory, California Institute
of Technology, under NASA contract 1407. Also based on data collected at:
the Subaru Telescope, which is operated by
the National Astronomical Observatory of Japan; the XMM-Newton, an ESA
science mission with
instruments and contributions directly funded by ESA Member States and
NASA; the European Southern Observatory under Large Program 175.A-0839,
Chile; Kitt Peak National Observatory, Cerro Tololo Inter-American
Observatory and the National Optical Astronomy Observatory, which are
operated by the Association of Universities for Research in Astronomy, Inc.
(AURA) under cooperative agreement with the National Science Foundation;
and the Canada-France-Hawaii Telescope with MegaPrime/MegaCam operated as a
joint project by the CFHT Corporation, CEA/DAPNIA, the NRC and CADC of
Canada, the CNRS of France, TERAPIX and the Univ. of
Hawaii.}
\altaffiltext{1}{California Institute of Technology, MC 105-24, 1200 East
California Boulevard, Pasadena, CA 91125.}
\altaffiltext{2}{Max Planck Institut f\"ur extraterrestrische Physik,
       Giessenbachstrasse 1, 85748 Garching, Germany.}
\altaffiltext{3}{Institute for Astronomy, University of  Hawaii, 2680 Woodlawn Drive,
Honolulu, HI, 96822 USA}  
\altaffiltext{4}{INAF--Osservatorio Astronomico di Bologna, via Ranzani 1,
I--40127 Bologna, Italy.}
\altaffiltext{5}{CFHT corporation, 65-1238 Mamalahoa Hwy, Kamuela, HI, Ê96743}
\altaffiltext{6}{CEA/DSM-CNRS, Universite' Paris Diderot, DAPNIA/SAp,
Orme des Merisiers, 91191, Gif-sur-Yvette, France.}
\altaffiltext{7}{Department of Physics and Astronomy, University College London, Gower Street, London, WC1E 6BT, UK}
\altaffiltext{8}{Steward Observatory, University of Arizona, Tucson, AZ 85721.}
\altaffiltext{9}{Department of Physics, Eidgenossiche Technische Hochschule
  (ETH), CH-8093 Zurich, Switzerland.}
\altaffiltext{10}{European Southern Observatory, Karl-Schwarzschild-str. 2,
  85748 Garching, Germany.}
\altaffiltext{11}{Institut d'Astrophysique de Paris, UMR 7095, CNRS, Universit\'e Pierre et Marie Curie, 98 bis Boulevard Arago, F-75014 Paris, France.}
\altaffiltext{12}{Riverside.}
\altaffiltext{13}{Physics Department, Graduate School of Science and Engineering, Ehime University, 2-5 Bunkyo-cho, Matsuyama 790-8577, Japan.}
\altaffiltext{14}{LBT Observatory, University of Arizona, 933 North Cherry Avenue, Tucson, AZ 85721-0065.}
\altaffiltext{15}{Physics Department, Brown University, Box 1843, Providence, RI 02912.}
\altaffiltext{16}{Observatoire Astronomique de Marseille-Provence, ÊPole de l'etoile site de Chateau-Gombert, 38,  rue Frederic Joliot-Curie 13388 Marseille Cedex 13, France}
\altaffiltext{17}{Research Center for Space and Cosmic Evolution, Ehime University, 
        Bunkyo-cho 2-5, Matsuyama 790-8577, Japan}
\altaffiltext{18}{Istituto Nazionale di Astrofisica (INAF) - Osservatorio Astronomico di Bologna, via Ranzani 1, 40127 Bologna, Italy}
\altaffiltext{18}{Department of Astronomy, Columbia University, MC2457, 550 W. 120 St. New York, NY 10027}
\altaffiltext{19}{Spitzer Science Center, California Institute of Technology, Pasadena, CA 91125}
\altaffiltext{20}{Max Planck Institut f\"ur Astronomie, K\"onigstuhl 17, D-69117 Heidelberg, Germany}
\altaffiltext{21}{Harvard-Smithsonian Center for Astrophysics 60 Garden St., Cambridge, Massachusetts 02138 USA}
\altaffiltext{22}{Carnegie Observatories, 813 Santa Barbara Street, Pasadena, California, 91101 USA}
\altaffiltext{23}{Dept of Physics, Yale University, PO Box 208121, New Haven, CT 06520-8121}
\altaffiltext{24}{Astronomical Institute, Graduate School of Science, Tohoku University, Aramaki, Aoba, Sendai 980-8578, Japan}

  

\begin{abstract}
We present photometric  redshifts and spectral energy  distribution (SED) classifications for  a sample  of 1542 optically identified sources detected with {\it XMM} in the COSMOS field. Our template fitting classifies 46 sources as stars and 464 as non-active galaxies, while the remaining 1032 require templates with  an AGN contribution. High  accuracy   in  the  derived photometric redshifts was accomplished as the result of  1) photometry in  up to 30  bands with high significance  detections, 2) a  new set of  SED  templates   including 18   hybrids covering  the   far-UV  to mid-infrared, which have been  constructed by the combination of AGN and non-active  galaxies templates, and  3) multi-epoch observations that have been  used to correct for variability  (most important for type  1  AGN).  The  reliability  of  the  photometric redshifts  is evaluated  using the  sub-sample   of  442  sources  with  measured spectroscopic  redshifts.  We achieved an accuracy  of $\sigma_{\Delta   z/(1+z_{spec})}   =  0.014$ for i$_{AB}^*<$22.5 ($\sigma_{\Delta z/(1+z_{spec})} \sim0.015$  for  i$_{AB}^*<$24.5). The high accuracies  were accomplished for both type 2  (where the SED  is often dominated  by the host  galaxy) and type 1 AGN and QSOs out to $z=4.5$. The number of outliers  is a large improvement over  previous photometric redshift estimates for X-ray selected sources (4.0\% and 4.8\% outliers for i$_{AB}^*<$22.5 and i$_{AB}^*<$24.5, respectively). We show that the intermediate band photometry is vital to achieving accurate photometric redshifts for AGN, whereas the broad SED coverage provided by mid infrared ({\it Spitzer}/IRAC) bands is important to reduce the number of outliers for normal galaxies.
 \end{abstract}




\section{Introduction}

It is now well established that  AGN play a significant role in cosmic
evolution -- tracing the growth  of super massive black holes in galaxy
nuclei and providing important energetic feedback to the host galaxies
and  the  IGM.  Characterization  of  the  AGN  properties  and  their
redshifts  is  therefore  of  paramount  importance  to  understanding
cosmological evolution.   The mass of the central  super massive black
hole  (SMBH)  directly  correlates  with  the mass  and  the  velocity
dispersion     of     the     bulge     of     its     host     galaxy
\citep{Gebhardt:2000jt,Ferrarese:2000hb}    and   these   correlations
apparently persist  both for normal, less-active galaxies  and for AGN
\citep{Mclure:2002la}. This  is strongly suggesting  a co-evolution of
galaxies and their nuclear BHs.  Unfortunately, given the low fraction
of galaxies with high luminosity AGN and the fact that those with high
luminosity may give a biased view of the overall importance of the AGN
feedback,  the co-dependent  evolution  of galaxies  and  BH has  been
pursued  very little  at high  redshift,  where the  majority of  both
galaxy  and AGN  evolution occurs.  Here  we address  this problem  --
attempting to  identify a much  larger sample of  AGN over a  range of
luminosity  and  to provide  classifications  and  redshifts for  much
fainter samples.
 
At present, no criterion exists to identify all AGN in a field using a
single photometric  band. Optical,  infrared, radio, narrow  bands and
X-ray  techniques   each  yield  incomplete,   partially  overlapping,
sub-samples of the overall AGN  population.  A complete census can only
be  achieved  by  combining  the  results  of  the  various  selection
techniques (Zamorani et~al.,  in preparation). Nevertheless, the X-ray
band  appears to  be  very  efficient and  rather  complete for  most
classes of AGN \citep{Brandt:2005fj}.

The COSMOS  survey provides full multi-wavelength  data-sets from X-ray
to  radio and  we focus  here  on {\it  XMM}--COSMOS selected  sources
\citep{Hasinger:2007dn} using spectral  energy distributions (SEDs) to
disentangle  between  non-active  galaxies  and  AGN  and  to  provide
accurate  photometric  redshifts.   Photometric redshifts  for  normal
(inactive) galaxies can reach  an accuracy of $\sigma_{\Delta z/(1+z)}
$    =    0.05    \citep{Grazian:2006kx,Ilbert:2006vl}    or    better
\citep{Gabasch:2004qq,Wolf:2004yq} and most  recently better than 0.02
(Ilbert et~al. 2008).  In contrast, for  AGN the most
accurate          results         have          been         0.06--0.1
\citep{Zheng:2004gd,Polletta:2007cs,Rowan-Robinson:2008ph}          for
redshift z$<$1.

Uncertainties in the  SED fitting of AGN come  from two major sources.
First,  AGN can vary  with time  and quasi-simultaneous  photometry is
required to  obtain a snapshot  SED. Second, the observed  emission is
often  a  superposition of  AGN  and  host  galaxy. Depending  on  the
properties  of the host  and the  relative power  of the  nucleus, the
different  spectral  bands can  be  dominated  by  either of  the  two
components.  For example, NGC3079 is a Seyfert 2 galaxy, identified on
the  basis of  its  optical nucleus  and  a radio  jet,  yet its  {\it
  Spitzer}/IRS spectrum  is often used  as a template for  non active,
starburst  galaxies in  the infrared  \citep{Weedman:2006ve}.   On the
other  hand, Mrk231, a  Seyfert 1  and ultra-luminous  infrared galaxy
(ULIRG), exhibits  strong silicate absorption in the  IRS spectrum and
is therefore highly obscured  with much AGN (and starburst) luminosity
re-emitted in the far infrared.  Clearly, to analyze samples which, as
{\it XMM}-COSMOS, contain both galaxies  and AGN, one needs: 1) a fine
sampling and full spectral  coverage with photometry spanning from the
UV  to  mid-infrared and  2)  spectral  energy distribution  templates
representing the  full range  of AGN and  QSOs together  with variable
mixing fractions of AGN and host galaxy emission.

Only recently, with  {\it Spitzer} and {\it GALEX}  providing full SED
templates      of      AGN      beyond     the      optical      bands
\citep[e.g.][]{Polletta:2007cs} such  analysis has become  possible at
moderately high redshift.

In this paper we derive photometric redshifts for the {\it XMM}-COSMOS
sources -- mostly dominated  by an AGN (see $\S$~\ref{sec:results}) --
achieving  an  accuracy  comparable   that  obtained  for  non  active
galaxies. The  COSMOS field  \citep{Scoville:2007uo} with its  size (2
square degrees) and deep photometry  in 30$^+$ bands from the radio to
the X-ray, is an excellent testbed for such a census.
  
 In $\S$~\ref{sec:dataset}, we describe the sample and the photometric
 data set.  The variability analysis is presented 
 $\S$~\ref{sec:variability}.  In $\S$~\ref{sec:fitting} the procedure
 used for the AGN classification and photometric redshift
 determination is discussed.  The results are given in
 $\S$~\ref{sec:results} and the description on the released catalog
 are described in $\S$~\ref{sec:catalog}.  The discussion and the
 conclusions end the paper in $\S$~\ref{sec:discussion} and
 $\S$~\ref{sec:conclusions}.  Throughout this work we use AB
 magnitudes and assume H$_0$=70\,kms$^{-1}$Mpc$^{-1}$,
 $\Omega_{\Lambda}$=0.7, and $\Omega_{M}$=0.3.

\section{The data set}
\label{sec:dataset}
\subsection{The {\it XMM}-COSMOS sample}
\label{sample}
The COSMOS field  has been observed with XMM-{\it  Newton} for a total
of                   $\sim                   1.55$                  Ms
\citep{Hasinger:2007dn,Cappelluti:2007fj,Cappelluti:2008}    at    the
homogeneous  depth  of  $\sim  50$  ks. The  final  catalogue  includes
point--like sources  detected above a  given threshold with  a maximum
likelihood detection  algorithm in  at least one  of the  soft (0.5--2
keV),  hard  (2--10 keV)  or  ultra-hard  (5--10  keV) bands  down  to
limiting fluxes of 5$\times 10^{-16}$, 3$\times 10^{-15}$ and 5$\times
10^{-15}$  erg   cm$^{-2}$  s$^{-1}$,  respectively   (see  Cappelluti
et~al. 2007, 2008 for more  details).  The adopted threshold ($Lik > $
10) corresponds to a probability of $\sim 4.5\times10^{-5}$ that a source
is a background fluctuation.  A detailed X--ray to optical association
using COSMOS  optical, near-  and mid-infrared catalogs  was performed
using       the      Maximum      Likelihood       (ML)      technique
\citep{Brusa:2007fp,Brusa:2008}.   For   the  sub-field  of   the  {\it
  XMM}-COSMOS   covered   also   by   {\it   Chandra},   the   optical
identifications  have  been  augmented  with the  more  accurate  {\it
  Chandra} positions  (Civano et~al. 2008;  Elvis et~al. 2008).   As a
result, the sample can be divided into the following subgroups:
\begin{itemize}

\item   1528  sources   (78.9\%)  have   unique  and   secure  optical
  counterparts  within 5.\arcsec0  from the  {\it XMM}  coordinates. We
  will refer to these sources as the "secure sample".
\item 52  sources are  brighter than i$^*$=16.5  and are  saturated in
  many (if not  all) optical bands.  37 of these are associated with
  bright  stars, and  1 is  a  galaxy with  spectroscopic redshift  at
  0.122. These  38 sources are removed  from the input  catalog as all
  the  photometry is  saturated.  We  will refer  to the  remaining 14
  sources as the "bright sample".
\item  10  sources are  undetected  in  the  optical catalog  and  the
  identification  of  the  counterpart   is  done  in  the  near-  and
  mid-infrared   using   the  K-band   (McCracken   et~al.  2008,   in
  preparation)                         and                        IRAC
  catalogs\footnote{http://irsa.ipac.caltech.edu/data/COSMOS/}.     For
  this  "optically  undetected sample"  not  more  than 5  photometric
  points are available.
  \item     233     sources      have     an     ambiguous     optical
    cross-identification. For these  objects, the ML technique yielded
    2 sources with similar probability.  This sample is referred to as
    having "ambiguous counterparts".
\end{itemize}
The  photometric  redshift  determination  is  tuned  using  only  the
"secure" and the "bright" samples  (1528+14 =1542 sources); all of the
results and  the discussion below  refer only to them.   The optically
undetected     sample    will     be    discussed     separately    in
$\S$~\ref{sec:faint},   while   for    the   sample   with   ambiguous
counterparts, we will provide photometric redshift for the primary and
secondary counterpart in the catalog  but they are not included in the
redshift distribution or in the discussion.
   
\subsection{Multi-wavelength data} 
   
For  the 1542 {\it  XMM} sources  associated with  an optical/infrared
counterpart, the  photometric redshift  has been computed  using broad
and  intermediate   band  photometry  in  the   range  1350\AA$<\lambda<$8$\mu$m
~\  using  {\it  GALEX},  Subaru, SDSS,  CFHT  and  {\it
  Spitzer}/IRAC.

 The optical/near -infrared photometry  is extracted from the recently
 updated COSMOS  photometric catalog (Capak et~al.  2008). Compared to
 the  previous   catalog  \citep{Capak:2007db},  the   new  photometry
 implements:  1) a more  accurate algorithm  for source  detection, 2)
 new, deeper  (and with better  seeing) u$^*$- and  K$_S$-band imaging
 from  CFHT  and  3)  new   J-band  data  (UKIRT)  and  12  additional
 intermediate and  2 narrow bands (from Subaru;  Taniguchi et~al 2007,
 Taniguchi et~al. 2008, in  preparation).  The fluxes were measured in
 fixed apertures of 3.\arcsec0  diameter, on PSF--matched images (FWHM
 of 1.\arcsec5). Simulations  \citep{Capak:2007db} show that the flux
 in 3.\arcsec0 aperture  corresponds to 0.759 of the  total flux for a
 point-like source.
 
For  each  optical/infrared  counterpart  of  a  source  in  the  {\it
  XMM}-COSMOS  sample,  the  closest   {\it  GALEX}  source  from  the
de-blended,     PSF-fitted    {\it     GALEX}-COSMOS     catalog    of
\cite{Zamojski:2007lq} was considered  for inclusion in the photometry
data set. Only 625 sources do have a GALEX counterpart.
 
IRAC photometry  was also extracted from the  IRAC-COSMOS catalog.  In
this case, the possible IRAC  counterpart of each {\it XMM} source was
taken to be  the closest IRAC source within  1.\arcsec0 of the optical
counterpart.  For  approximately 5\% of the  {\it XMM}-COSMOS sources,
blending  of multiple  IRAC sources  makes it  impossible  to reliably
extract a correct IRAC flux.  Additional 5\% fall in masked regions in
the IRAC images and therefore lack mid-infrared photometry.
 
The  3.\arcsec8  aperture fluxes  given  in  the  COSMOS IRAC  catalog
(aperture 2 in  the catalog) where then translated  to total fluxes by
division  by factors  of  0.765, 0.740,  0.625,  0.580 for  3.6$\mu$m,
4.5$\mu$m, 5.8$\mu$m,  and 8$\mu$m, respectively (see  the readme file
associated to the catalog and Surace et~al. 2004 for details).

Lastly, we multiply the total flux from {\it GALEX} and IRAC by 0.759,
in order to  re-scale the photometry to the  optical and near-infrared
one.  All photometry has been  transformed to the AB magnitude system.
The  available bands,  depth and  observing epochs  are  summarized in
Table~\ref{tab:runs}.

\section{Variability analysis}
\label{sec:variability}
AGN exhibit non-periodic flux variations  on time scales of minutes to
decades  \citep[see][for  a  review]{Wold:2007ad}.   Thus,  photometry
collected non-simultaneously  will be  affected by variability  and 
may  be not  representative of  a  snapshot SED.   Wolf et~al.  (2004)
pointed out  that ignoring variability can  significantly increase the
errors in photometric redshift estimations of AGN.

The  COSMOS  optical  photometry  has  been acquired  in  five  epochs
distributed    over     $\sim$6    years    (see     column    7    in
Table~\ref{tab:runs}).  Within each epoch  (apart from  2005), several
individual  filter  observations were  distributed  over  less than  3
months,  covering  the  whole  optical range.  The  
photometry in each filter has generally a very high calibration accuracy. This allows
to study and  correct AGN time variability over  time scales of years,
while shorter variability can not easily be addressed.
\begin{table*}[htdp]
\caption{Photometric coverage and depth of  the COSMOS data-set used for the photometric redshift estimation.\label{tab:runs}}
\begin{center}
\begin{tabular}{c|c|c|c|c|c|c}
 Filter & telescope &effective $\lambda$ & FWHM &zp$_{corr}$ &Depth&Time of observations\\
        &     &$\AA$              & $\AA$  &mag$_{AB}$&  mag$_{AB}$& UTC          \\
 \hline
 \hline        
u$^*$    &CFHT     &    3911.0  &    538  &  0.05   &26.50  &  Jan 2003 - Apr 2007 \\
 B$_J$  &Subaru    &    4439.6  &    807   & -0.24    &27.00  &  Jan 2004 \\
V$_J$   &Subaru    &    5448.9  &    935  & -0.09   &26.60  &        Feb 2004 \\
 g$^+$   &Subaru  &    4728.3 &   1163   & 0.02     &27.00  &       Jan 2005 \\
r$^+$     &Subaru &    6231.8 &   1349  & 0.00    &26.80  &        Jan 2004 \\
i$^+$      &Subaru   &    7629.1  &   1489  & 0.02    &26.20  &       Jan 2004 \\
 z$^+$    &  Subaru &  9021.6 &    955   & -0.04  &25.20  &       Jan 2004\\
J         &    UKIRT    & 12444.1 &   1558   & 0.12    &23.70  &        Mar 2006 \\
K$_S$   &  CHFT     &    21480.2 &   3250   & -0.05   &23.70  &       Mar 2007 \\
i$^*$   &    CHFT         &    7629  &   1460  & -0.01  &24.00  &       Jan 2004 \\   
u       &      SDSS        &    3623  &    557   & 0.00     &22.00  &         Jul 2001 \\
g       &      SDSS         &    4677  &   1171   & 0.00    &22.20  &       Jul 2001 \\
r       &        SDSS         &    6159  &   1123  & 0.00   &22.00  &      Jul 2001 \\
i       &         SDSS         &    7480  &   1291   & 0.00     &21.30  &     Jul 2001 \\
z      &         SDSS      &    8884  &    943  & 0.00    &20.50  &        Jul 2001 \\
IA427  &Subaru &    4256.3 &    207   & 0.04    &25.82  &        Jan 2006 \\
IA464  &Subaru   &    4633.3  &    218   & 0.01     &25.65  &        Feb 2006 \\
IA484   &Subaru  &    4845.9 &    229   & 0.00     &25.60  &        Jan 2007 \\
IA505    &Subaru &    5060.7  &    231   & 0.00    &25.55  &        Feb 2006 \\
IA527    &Subaru &    5258.9  &    242  & 0.03    &25.62  &        Jan 2007 \\
IA574    &Subaru &    5762.1  &    272  & 0.08    &25.61  &      Jan 2007 \\
IA624    &Subaru &    6230.0  &    301   & 0.00     &25.60  &        Dec 2006 \\
IA679    &Subaru &    6778.8  &    336   & -0.18    &25.60  &         Feb 2006 \\
IA709    &Subaru &    7070.7  &    316  & -0.02    &25.65  &         Feb 2007 \\
IA738    &Subaru &    7358.7  &    324   & 0.02     &25.60  &          Jan 2007 \\
IA767    &Subaru &    7681.2  &    364  & 0.04    &25.60  &        Mar 2007 \\
IA827    &Subaru &    8240.9  &    344 & -0.02   &25.39  &      Jan 2006 \\
IRAC1   &{\it Spitzer} &    35262.5&    7412  & 0.00     & 23.90 &       Jan 2006 \\
IRAC2   &{\it Spitzer}  &    44606.7 &   10113  & 0.00     & 23.30 &       Jan 2006 \\
IRAC3    &{\it Spitzer} &    56764.4 &   13499  & 0.01   & 21.30 &        Jan 2006 \\
IRAC4    &{\it Spitzer} &    77030.1 &   28397  & -0.17    & 21.00 &         Jan 2006 \\
FUV      &{\it GALEX}&     1551.3 &     231  & 0.31     &25.69  &         Feb 2004 \\
NUV      &{\it GALEX}&     2306.5 &     789  & -0.02   &25.99  &         Feb 2004 \\

\end{tabular}
\end{center}
\end{table*}

Comparing  the  extracted fluxes  at  different  epochs,  a number  of
objects  show  clear  time  variability.  As  an  example,  in  Figure
\ref{fig:variability} we  show the  variability of source  XID 17  and the
correction  that  we  apply.  In  the  upper  panel  the  raw  optical
photometry of this source is shown as black data points with different
symbols for each epoch. There is obviously a large scatter between the
photometry  at  different  epochs.  The  time  variability  correction
proceeded in the following way:
\begin{figure}[h]
   \centering
   \includegraphics[scale=0.4]{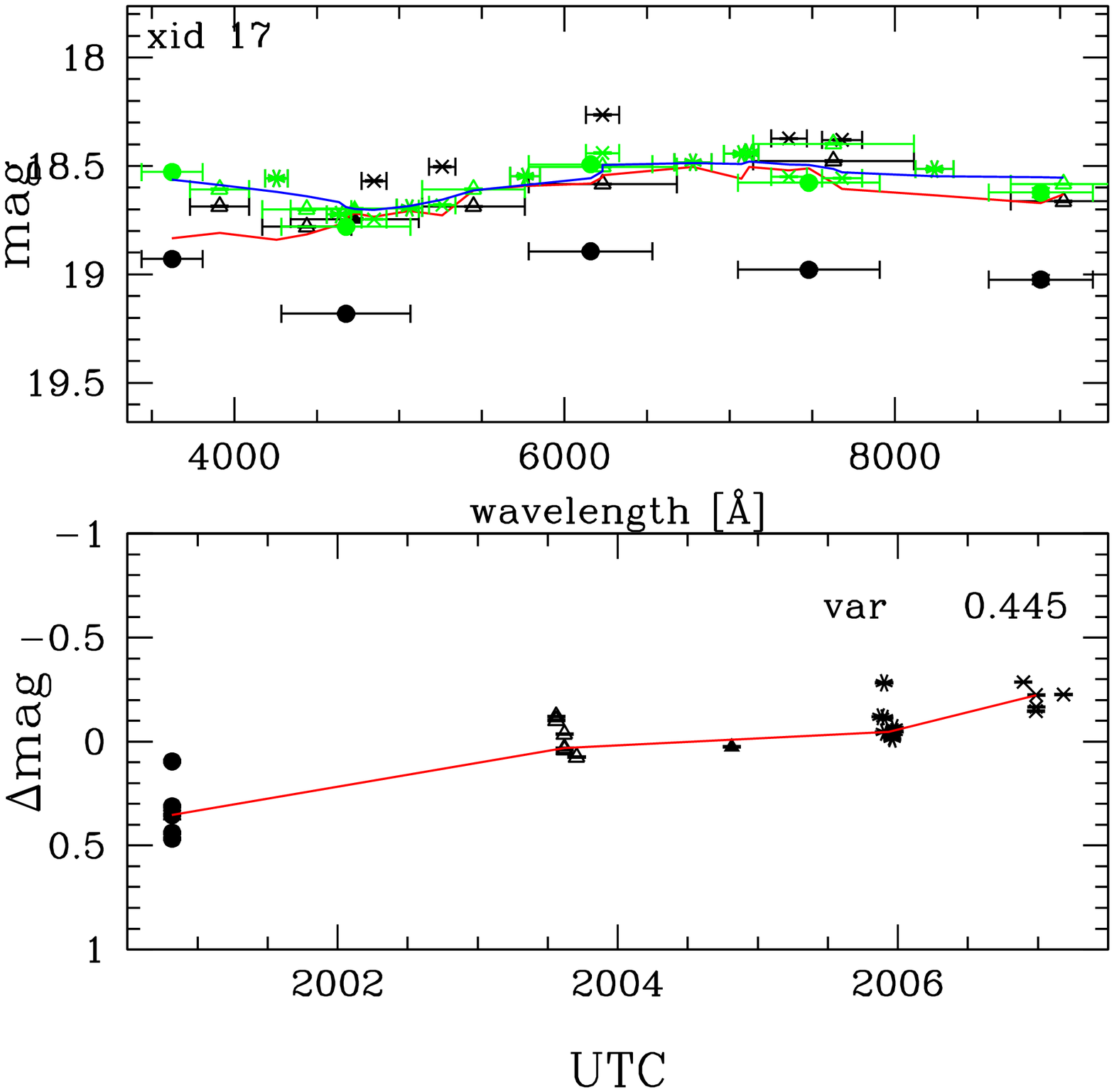} 
   \caption{Time variability for  the source with XID 17. Bottom:  The source was observed in 5 epochs, each epoch marked with a different symbol. 
 The 5 groups are, from left to right: SDSS,  Subaru broad band images (Subaru$_{BB}$), CFHT i band,  the first set of intermediate  bands Subaru images (Subaru$_{IB1}$)  and the second epoch of intermediate  bands Subaru images (Subaru$_{IB2}$). The optical photometry  varied  up to 0.6 mag  over the  6 years  of COSMOS data acquisition. The red line connects the median values for each group of observations. Top: the photometric points of the source before (black) and after (green) our correction. The red and blue lines represent the smoothed spectral shape of the object before and after the correction for variability (see text for more details).}
\label{fig:variability}
\end{figure}

\begin{itemize}

\item  In order to  flatten out  the overall  spectral shape  to first
  order,  the  wavelength dependence  of  the  optical magnitudes  was
  simply  smoothed by  a running  average filter.  The result  of this
  smoothing  is shown  as the  red solid  line in  the upper  panel of
  Figure \ref{fig:variability}.

\item  This smoothed  spectral  shape was  then  subtracted from  each
  individual data  point in order to calculate  the relative magnitude
  differences    shown    in     the    lower    panel    of    Figure
  \ref{fig:variability}.  Here  the  magnitude differences  $\Delta$m  are
  plotted against the observation date  of the 5 separate epochs, $n$,
  of contemporaneous observations. For a constant source all points in
  this  panel should  be within  the measurement  uncertainties around
  $\Delta$m=0.  In   this  particular  example,  the   source  XID  17
  apparently brightened by more than 0.6 magnitude over the 6 years.

\item For the  epochs $n$=1,2,4 and 5 the  median $\Delta$m$_n$ of the
  magnitude differences  of all optical  bands observed in  this epoch
  was calculated.  For epoch  $n$=3, which consists  of only  one data
  point, $\Delta$m$_3$ was computed  by a linear interpolation between
  $n$=2 and $n$=4.

\item The  correction offset for a  given epoch $n$  is the difference
  $offset_n=\Delta{m_n}-\Delta{m_{n=4}}$, where $n$=4  is the epoch of
  the Subaru$_{IB1}$ and IRAC observations (Dec 2005--Mar 2006).  Each
  point in  the plot represents the  deviation of the  photometry in a
  given filter observed in epoch $n$, from m$_{n}$.

\item In order to derive  the closest approximation to a snapshot SED,
  we  corrected for  variability subtracting  the value  of $offset_n$
  from each  of the multi-epoch  photometry data points,  resulting in
  the    green    data    points     in    the    upper    panel    of
  Figure~\ref{fig:variability}.

\item Similar to the red curve,  the thin blue solid line in the upper
  panel of Figure~\ref{fig:variability}  shows the smoothed spectral shape
  of the variability corrected magnitudes.

\end{itemize}
In  the case  of  source  XID 17  the  variability correction  clearly
resulted in a  significant reduction of the scatter  in the photometry
and thus an improvement  in the photometric redshift determination. As
discussed  above, this  correction  scheme only  works  for slow  time
variability  and there are  examples of  sources, where  a significant
time variability could  not be reduced by this  method. However, given
the temporal  sampling of the diverse  data sets, it is  the best that
could be  done and this  correction was found to  significantly reduce
the  number   of  outliers  for  the   spectroscopic  sub-sample  (see
$\S$~\ref{sec:discussion}).

\begin{figure}[htbp]
   \centering
   \includegraphics[scale=0.4]{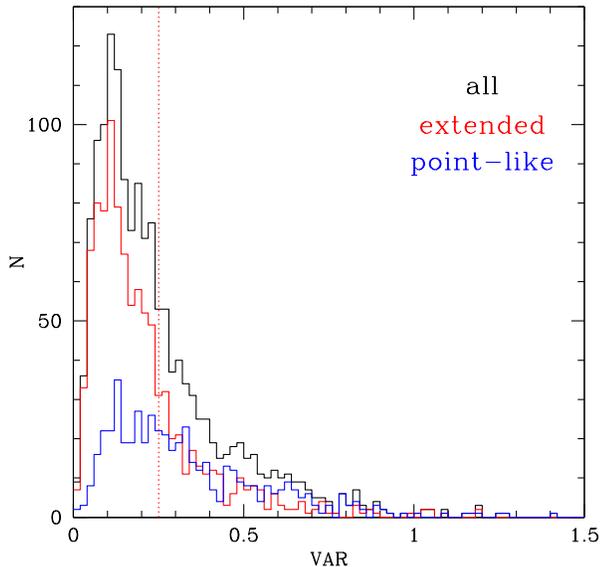} 
   \caption{ Histogram  of the parameter {\it VAR} for  all the sources (black solid line) extended (red)  and point-like sources (blue). The vertical dotted line at $VAR$=0.25 defines our adopted separation between host (low $VAR$) and AGN (high $VAR$) dominated sources.  }
\label{fig:var_histo}
\end{figure}

As anticipated from the known variability time scales of AGN, the {\it
  XMM}-COSMOS sources exhibited a variety of behaviors -- some sources
evolved gradually while others displayed more extreme outbursts. We
quantify the observed variability with the parameter,
\begin{equation}
VAR = \sqrt{\sum_{n=1,2,3,5}(\Delta{m_n}-\Delta{m_4})^2.}
\end{equation}
In Figure \ref{fig:var_histo} the distribution of {\it VAR } for all the
sources (black solid line) is shown.
 
\section{Fitting technique}
\label{sec:fitting}
To derive the intrinsic SEDs and redshifts of the observed sources, we
used    the     publicly    available    LePhare     code    (Arnouts,
Ilbert)\footnote{http://www.oamp.fr/~people/arnouts/LE\_PHARE.html}.  For  this  fitting,  a  library  of
expected  intrinsic  rest-frame  SEDs  of  the  source  population is 
supplied and  a $\chi^2$  minimization is used  to solve for  the most
likely redshift,  SED type  and intrinsic extinction.  A user-supplied
prior  can   also  be   used.  In  the   following  we   describe  the
transformation of the template SEDs to the observed photometric bands,
the selection  of template  SEDs, the adopted  extinction law  and the
control of source variability which is important for AGN.

 The accuracy of the photometric redshift fitting was evaluated
  using a sub-sample of 396 {\it XMM} sources with i$^*\le$22.5 and
  secure spectroscopic redshift.  90 sources had optical spectroscopy
from Sloan Digital Sky Survey
\citep{York:2000rz,Adelman-McCarthy:2008yq} while 306 were observed
within the COSMOS collaboration (235 observed with IMACS and MMT
\citep{Trump:2007fv,Prescott:2006ek}; 71 observed by $z$COSMOS with
VIMOS \citep{Lilly:2007qr}).  The spectroscopic sample was used to
train/evaluate the fitting procedure and the accuracy quantified as
$\sigma_{\Delta z/(1+z_{spec})} = 1.48 \times {\rm median} (|z_{\rm
  phot}-z_{\rm spec}|/(1+z_{\rm spec})$ \citep{Hoaglin:1983bh}, where
$z_{\rm phot}$ and $z_{\rm spec}$ are the photometric and
spectroscopic redshifts, respectively.  As the reliability of
photometric redshifts trained on local/bright spectroscopic samples
may not hold for low-~luminosity or high redshift sources for which
the SED templates may not longer be representative, we have also
performed an {\it a posteriori} test using spectroscopy which became
available only recently for 46 optically faint, K-band and {\it
  Spitzer}/MIPS 24$\mu$m selected sources (Lilly et~al. 2008;
Kartaltepe et~al. 2008).  These additional spectroscopic redshifts
indicated a high success rate with correctly recovered redshift also
for these faint sources (see $\S$~\ref{sec:check_faint}).

\subsection{Transformation of Template SEDs to Observed Photometric System}
To reduce the spectra template to the same  photometric system of the
observed  sources, all SEDs were  convolved with  the filter
transmission curves, which also  include the instrument efficiency. The
same  zero-points  derived for  the  new  COSMOS photometric  redshift
catalog of non X-ray detected galaxies (Ilbert et~al. 2008)
were used  for the {\it  XMM} sources. This zero-point  calibration is
extensively explained in  \cite{Ilbert:2006vl} and reported in column
5  of  Tab.~\ref{tab:runs}.  The  convolved  template  SEDs were  then
redshifted  in  $\Delta$z= 0.02  wide  bins  from  z=0.02 to  z=8  and
parabolically interpolated  for higher redshift  accuracy. The opacity
of the  inter-galactic medium was  taken into account as  described by
\cite{Madau:1995kx}.

\subsection{Galaxy and AGN Templates}
\label{sec:templates}
Ideally, a template library should represent all possible SEDs.
However, the larger the number of templates, the higher the risk of
degeneracy, which then can lead to an increasing number of
catastrophic failures.  To identify a suitable library, we run LePhare
on the spectroscopic sample of 396 sources with i$^*_{AB}<$22.5.
Here, we fixed the redshifts to the spectroscopic values and let the
code select the best fitting SED template.  Initially, the entire set
of 28 templates from
SWIRE\footnote{http://www.iasf-milano.inaf.it/~polletta/templates/swire\_templates.html},
was used, including elliptical and spiral galaxies, starbursts, ULIRG,
low luminosity AGN and QSOs (QSO1, QSO2, TQSO1, BQSO1 etc; see
Polletta et~al. 2007 for details on these templates).  We also added
two composite low- and high-luminosity QSO templates from SDSS and one
template of a blue, star-forming galaxy from the library used by
Ilbert et~al.  (2008), for a total of 31 templates.

First, we trimmed this list by discarding all templates that were not
selected for any of the 396 sources in the spectroscopic sample.
  Next, we selected for each galaxy type the template which was most
  often used.  Accordingly, the S0 and the I22491 templates were
  singled out to represent passive galaxies and starburst/ULIRGs,
  respectively.  This process reduced the original number of SEDs
from 31 to 12.

For a  large number of objects, significant  discrepancies between the
photometry and templates occurred long-ward of 1\,$\mu$m. This suggests
that the SEDs are  not properly representing the mid-infrared emission
component. The SED templates are from spectra of nearby galaxies where
the nuclear  flux is  well sampled  but the host  galaxy is  not fully
included in slits or fibers.  In contrast, for all {\it
  XMM} sources at intermediate/high  redshift, their host galaxy sizes
are  comparable to  the  seeing achievable  with ground-based  optical
telescopes and  our photometric data should include  a larger fraction
of the host galaxy emission \citep{Szokoly:2004vn}. Thus, for the {\it
  XMM}  sources analyzed here,  the SED  templates should  very likely
have  a larger  contribution  from  the host  galaxy  relative to  the
nuclear emission component.

\begin{figure}[htbp]
\centering
\includegraphics[scale=0.4]{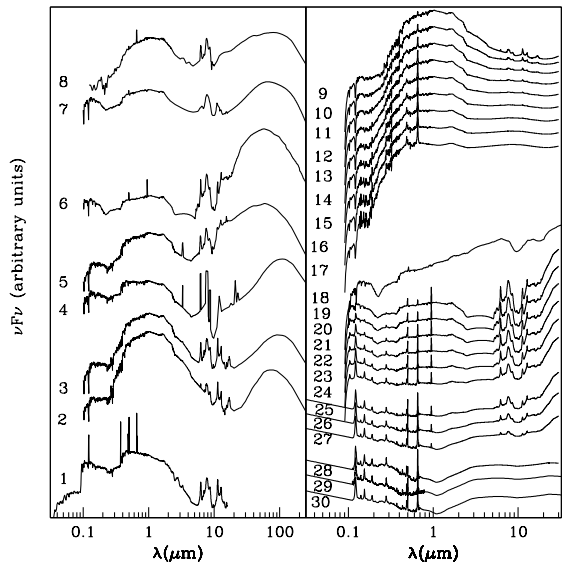} 
 \caption{ Visualization of the final choice of templates used in this paper. Numbers correspond to the IDs in Table~\ref{tab:SEDtab}.}
\label{fig:SEDfig}
\end{figure}

To better account  for the host galaxy component, for each
pair  of  AGN (type  1,  type  2, QSO1,  QSO2)  and  host galaxy  type
(elliptical,  spiral, starburst)  hybrid  templates were  constructed.
Here,  a pair  of  SEDs was  normalized  by the  integrated fluxes  in
different  pass  bands.  For  each  normalization,  nine hybrids  with
varying  ratio   (90:10,  80:20,  ...,  20:80,   10:90)  between  both
components were built.  After re-calculating the photometric redshifts
using these hybrid  SEDs, two sets of templates  provided the smallest
$\sigma_{\Delta  z/(1+z_{spec})}$  and number  of  outliers, $\eta$  (
defined      as     percentage      of      objects     for      which
$|z_{phot}-z_{spec}|>0.15(1+z_{spec})$.     These     sets     of     hybrids
are:\\  $\bullet$  starburst/ULIRG I22491  plus  TQSO1 templates,  both
normalized   between  5000\AA\,  and   5200\AA\,  (models   19--27  in
Table~\ref{tab:SEDtab} and in Figure~\ref{fig:SEDfig}); $\bullet$ S0 plus QSO2
templates, both normalized between  0.9$\mu$m and 1$\mu$m (models 9--17
in Table~\ref{tab:SEDtab} and in Figure~\ref{fig:SEDfig}).
\begin{table}[htdp]
\caption{Final template library used in this work.\label{tab:SEDtab}}
\begin{tabular}{c|c|c}
\hline
ID &Name\tablenotemark{1}& Type \\
\hline
\hline
 01& BC03&Blue starforming\tablenotemark{2}\\
 02&S0 &S0\tablenotemark{3} \\
 03&Sb &Spiral b\tablenotemark{3}\\
 04&Spi4 &Spiral c\tablenotemark{3}\\
 05&M82 &Starburst\tablenotemark{3}\\
 06&I22491 &Starburst/ULIRG\tablenotemark{3}\\
\hline
\hline
 07&Sey18 &Seyfert 1.8\tablenotemark{3}\\
 08&Sey2 &Seyfert 2\tablenotemark{3}\\ 
 09& S0-90\_QSO2-10 &Hybrid\tablenotemark{4} \\
 10& S0-80\_QSO2-20 & "\\
 11& S0-70\_QSO2-30 & "\\
 12&  S0-60\_QSO2-40& "\\
 13& S0-50\_QSO2-50 & "\\
 14& So-40\_QSO2-60 & "\\
 15& S0-30\_QSO2-70 & "\\
 16& S0-20\_QSO2-80 & "\\
 17& S0-10\_QSO2-90 & "\\
\hline
\hline
 18&Mrk231&Seyfert 1, BALQSO\tablenotemark{3}\\\
  19&I22491-90\_TQSO1-10 &Hybrid\tablenotemark{4} \\
  20&I22491-80\_TQSO1-20 & "\\
  21&I22491-70\_TQSO1-30 & "\\ 
  22&I22491-60\_TQSO1-40 & "\\
  23&I22491-50\_TQSO1-50 & "\\
  24&I22491-40\_TQSO1-60 &"\\
  25&pl\_I22491-30\_TQSO1-70& "\\
  26&pl\_I22491-20\_TQSO1-80 & "\\
 27&pl\_I22491-10\_TQSO1-90 &"\\
  28&pl\_QSOH &QSO high lum.\tablenotemark{5}\\
29&pl\_QSO & QSO low lum.\tablenotemark{6}\\
 30&pl\_TQSO1 & QSO  high IR lum.\tablenotemark{3} \\
  \hline
    \end{tabular}
    \tablenotetext{1}{Names are as in the original libraries.}    
  \tablenotetext{2}{Ilbert et~al. 2008}
 \tablenotetext{3}{\cite{Polletta:2007cs}}
 \tablenotetext{4}{This work}
\tablenotetext{5}{composite \#33 at http://www.sdss.org/dr5/algorithms/spectemplates}
\tablenotetext{6}{composite \#30 at http://www.sdss.org/dr5/algorithms/spectemplates}
 \end{table}

Lastly, a number of  sources showed discrepancies between the observed
and predicted flux  in the UV. This is likely  the result of intrinsic
absorption  in the  empirical templates.   Thus, we  extended  the QSO
dominated templates  25--30 (templates with  prefix "pl") into  the UV
using  a  power  law  component  with  spectral  index  $\alpha=-0.56$
\citep{Scott:2004zp}. Our final library includes 30 SEDs.
  
\subsection{Intrinsic Extinction}
To allow for the intrinsic reddening, two extinction laws were tested:
the SMC extinction derived by Prevot et~al. (1984) and the starburst
galaxy extinction law obtained by Calzetti et~al. (2000). We explored
E(B-V) values from 0 to 0.5 in steps of 0.05.  The two laws provide a
similar photometric redshift accuracy ($\sigma_{\Delta
  z/(1+z_{spec})}=0.014$), but with the SMC extinction law, the number
of outliers is reduced by 1.1\%.  In principle, highly obscured AGN
might have A$_V >1$.  However, none of the fits required E(B-V) $>0.4$
and 66\% of the sources were fit with E(B-V)$\leq$ 0.05. Less than 20
sources required E(B-V)$>$0.35.  This result is perhaps not totally
surprising, since  many of our templates are empirical and thus already 
  include some extinction.

\subsection{Luminosity Prior}
\label{sec:priors}
Applying a luminosity prior during the SED fitting can be important to
avoid unlikely combinations of source type, luminosity and redshift
which can lead to wrong solutions for the photometric redshift.
Quasars are usually defined as point-like sources which have an
absolute magnitude of M$_B\leq-23$. For lower luminosity AGN, the host
galaxy contributes to the observed flux and the object may be
classified as extended in good seeing data or, even more, in HST data.
In the spectroscopic sample we did not find any point-like source with
an absolute magnitude M$_B$ fainter than $-20$ and thus we adopted
this value as the lower limit on the luminosity prior.  While it
  is known that low luminosity AGN may have M$_{B}$ fainter than our
  limit, to use a lower threshold of M$_B$=-19 (as used by Polletta
  et~al. 2007 and Rowan-Robinson et~al. 2008), would increase the
degeneracy between galaxy and AGN templates and thus the outliers from 6.3\% to 8.0\%.

Type  1  AGN and  QSO  can be  identified  first  by their  point-like
appearance in the image or by  the amount of variability as defined in
$\S$~\ref{sec:variability}.  For the sources detected in the ACS/FW814
filter and  brighter than FW814$_{AB}$=24 the  peak surface brightness
is  measured and  used  to separate  extended  and point-like  sources
\citep{Leauthaud:2007fj}.  For the  500 sources missing this parameter
the star/galaxy  separation computed on the best  seeing Subaru R-band
image has been used to  separate extended and point-like sources.  The
sub-sample of  sources with  an extended optical  counterpart includes
989 sources, while  the sub-sample of point like  sources includes 553
members.

We also applied the prior in luminosity to sources with high
variability ($\S$~\ref{sec:variability}).  For extended sources, the
variability parameter $VAR$ can additionally be used to disentangle
between host and AGN dominated sources.  As expected, the majority of
the extended sources (Figure~\ref{fig:var_histo}, red distribution)
exhibit only low levels of variability ($VAR<$0.25). However, a tail
of higher {\it VAR} is present for low redshift type 1 AGN, where
apparently both the host galaxy and the variability of the nucleus are
detected.  In order to identify the best threshold, we run the
photometric redshift code using a wide range of $VAR$ parameters above
which the luminosity prior was applied also to extended objects.  The
prior is always used for point-like sources.  The smallest number of
outliers was achieved using a separation at $VAR=0.25$. Accordingly,
we found 252 extended sources in the $XMM$-COSMOS sample with
$VAR>0.25$.

In the following  we will refer to the point-like  and to the variable
sources  (805)  as  "qsov"  sample.  To this  sample  we  applied  the
luminosity  prior.  All  the  remaining extended  sources  (737)  with
$VAR\le0.25$  form the  "extnv" sample.  No  prior is  applied to  the
latter sample.
 
\section{Results}
\label{sec:results}

\subsection{Photometric redshift accuracy}

\begin{figure}[htbp]
\begin{center}
\includegraphics[scale=0.4]{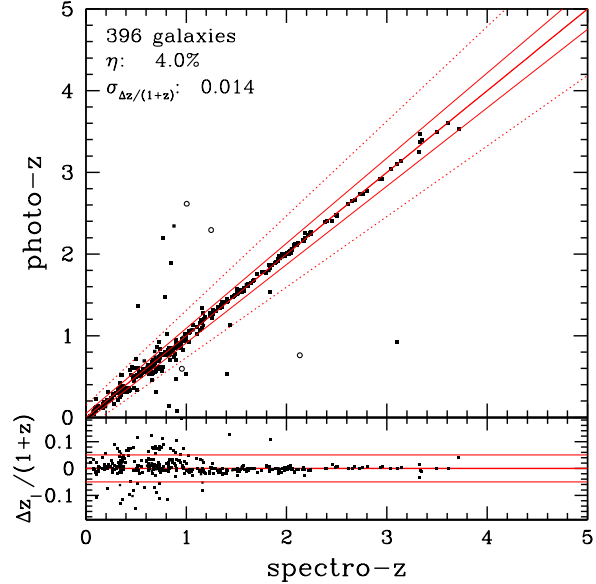}
\caption{Comparison of photometric and spectroscopic redshifts  for the  396  sources with i$^*_{AB}<$22.5. Empty circles represent sources  for which the second peak in the redshift probability distribution  agrees with the spectroscopic redshift. The lines correspond to 1) z$_{phot}=z_{spec}$ (thick solid),  2) z$_{phot}=z_{spec}\pm$0.05(1+z$_{spec}$) (solid)  and 3) z$_{phot}$=z$_{spec}\pm 0.15(1+z_{spec}$) (dotted).  }
\end{center}
\label{fig:zp_zs_bright}
\end{figure}

Following  the  procedure  described  in  the  previous  sections,  we
obtained photometric  redshifts for  all 1542 optical  counterparts of
the {\it XMM}-COSMOS  sample described in $\S$~\ref{sec:dataset}.  For
the  396 bright  spectroscopically  confirmed galaxies  we obtained  a
final   accuracy  of   $\sigma_{\Delta  z/(1+z_{spec})}   =0.014$  and
$\eta$=4.0\%  outliers  (see Figure~\ref{fig:zp_zs_bright}).   Considering
the  "qsov"  and  "extnv"   samples  separately,  the  accuracies  are
$\sigma_{\Delta     z/(1+z_{spec})}$=0.012     and     $\sigma_{\Delta
  z/(1+z_{spec})}$=0.019  while the  outliers  are $\eta$=6.3\,\%  and
$\eta$=2.3\,\%,    respectively     (see    the    first     row    of
Table~\ref{tab:summary}).

\begin{figure}[htbp]
\begin{center}
\includegraphics[scale=0.4]{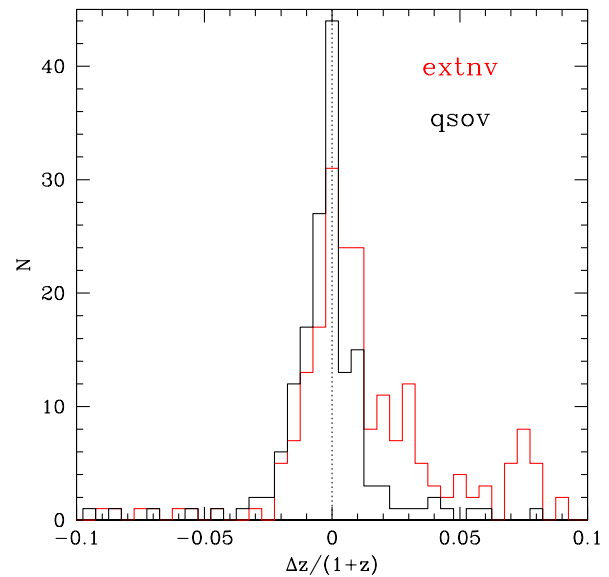} 
\caption{Histogram  of $\Delta$z/(1+z) for the ``extnv'' (red) and ``qsov''  (black) samples (see text for details). }
\end{center}
\label{fig:dz_histo}
\end{figure}

 Figure      \ref{fig:dz_histo}     shows     the      distribution     of
 $(z_{spec}-z_{phot})$/(1+z$_{spec}$)   for    the   "extnv"   and   "qsov"
 samples.  The  histogram  for  the  "extnv" sample  is  broader,  not
 perfectly  centered at  0 and  with  a small  asymmetric tail  toward
 positive  $\Delta z$  (i.e. z$_{phot}<z_{spec}$).  We found  that the
 scatter could be reduced for  the "extnv" sample by adding additional
 normal galaxy  SED templates, but this also  increases the degeneracy
 and the number of outliers.

 We were also able to characterize the nature of the small number of
 outliers and to understand the reasons why the redshifts were not
 correctly assigned.  First, 6 out of the 16 outliers (37.5\%) are
 blended with a nearby object, closer than 1.\arcsec5, while the
 spectroscopic follow up was centered on the right counterpart,
 identified on our ACS images.  In the entire {\it XMM}-COSMOS catalog
 we flagged 34 cases as potentially problematic, due to the presence
 of a close source.  Additionally, 2 of the remaining outliers miss
 photometry in the IRAC band (i.e. the source is either partially or
 completely in a masked area) and the remaining optical bands were not
 sufficient to find the correct redshift solutions.  Lastly, we found
 that for strongly varying sources ($VAR>0.25$) the available u$^*$
 fluxes were likely the problem -- the adopted u$^*$ magnitudes are an
 average between 4 runs of observations taken yearly between 2004 and
 2007. Adding 0.2 mag in quadrature to the measurement uncertainty
 associated with the u$^*$ band photometry for the outliers, 3 of the
 remaining 8 outliers would now have the correct redshift due to the
 reduced weighting of the u$^*$ band; an additional one would have the
 second comparable but lower peak of the redshift probability
 distribution at the right redshift.  We decided against
 systematically reducing the weighting of the u$^*$ band for the
 entire sample, as this would degrade the quality of the final result.

 \subsection{Redshift and Templates Distribution}
\label{sec:dissed}
\begin{figure}[htbp]
   \centering
   \includegraphics[scale=0.4]{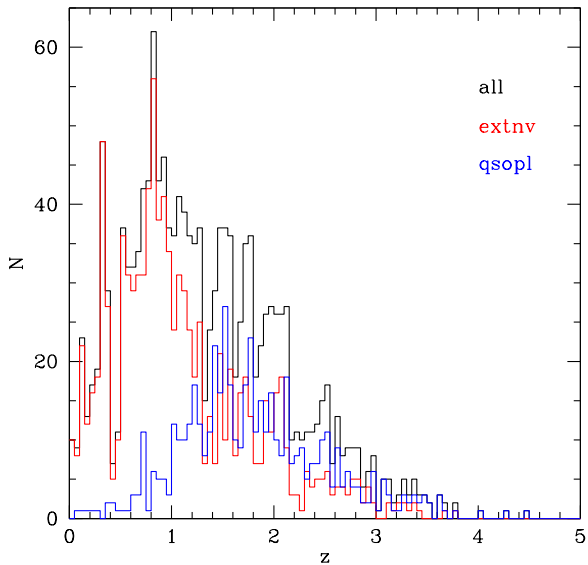} 
   \caption{ Photometric redshift distribution for the entire  sample (black line), ``extnv'' (red) and ``qsov'' (blue) samples.}
\label{fig:zphot_histo}
\end{figure}

Figure~\ref{fig:zphot_histo}  shows the photometric  redshift distribution
for the complete {\it XMM} sample.  As expected, the large majority of
the sources  at z $<$  1 are extended  and not strongly  time variable,
while  most of  the point-like  and varying  sources are at  z $>$  1. The
redshift   distribution  shows   significant  peaks   at  z$\sim$0.35,
$\sim$0.8,   corresponding  to   spectroscopically   confirmed  galaxy
over-densities          in          the          COSMOS          field
\citep{Scoville:2007kx,Finoguenov:2007fv,Lilly:2007qr,Gilli:2008}. This
suggests that the clustering of the normal galaxies and the AGN follow
each other.

\begin{figure}[h]
\centering
\includegraphics[scale=0.4]{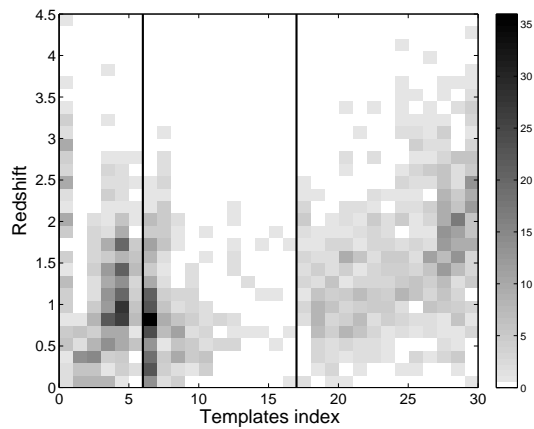} 
\caption{ Density map of SED templates distribution as a function of photometric redshift.}
\label{fig:densitymap}
\end{figure}

The distribution of the selected SED templates as a function of
redshift is shown in Figure~\ref{fig:densitymap}.  Templates of normal
galaxies (templates 1 to 6) or type 2 low luminosity AGN (e.g. Seyfert
1.8, template 7) are most frequent at z$<$2.  At higher redshift,
mainly QSO templates (templates 28 to 30) have been selected.  For
some of the lower redshift galaxies, the detected X-ray emission may,
in fact, arise from star forming regions rather than nuclear activity,
while at high redshift, the X-ray emission is predominantly from AGN.

In Section~\ref{sec:reliabilitySED}, we show that the best fit SED
templates are generally consistent with  other properties of the
host galaxies.  We note that Figure~\ref{fig:densitymap} clearly reveals
that templates 14 to 17, with an increasing contribution of the QSO2
SED, were rarely selected.  However, running the photometric redshift
fitting without these templates resulted in a significant increase in
outliers, indicating that for a few sources these templates are
essential.  In summary, we classify 464 sources as best fit by a
non-active galaxy (templates 1-6) while 1032 are best fit with a
template which include an AGN component (templates 7-30).

\subsection{Optically undetected sample}
\label{sec:faint}
As discussed in $\S$\ref{sec:dataset}, 10 XMM-COSMOS sources are not
detected in our optical broad or intermediate filters.  However,
unique counterparts for these sources have been identified in the
K-band and in the IRAC catalogs.  These are potentially very high
redshift sources (z$>3$; Koekemoer et~al. 2007, Brusa et~al. 2008), as
  suggested for similar MIPS 24 selected objects
  \citep{Houck:2005ys,Weedman:2006fr,Yan:2007zr,Sajina:2007mz}. However
  doubts on this interpretation has recently been presented by
  \cite{Brand:2008ly} who observed 16 X-ray loud, optically faint
  sources with {\it Spitzer}/IRS and measured redshifts z$\sim$2 for
  all targets.

\begin{figure*}[htbp]
\begin{center}
\includegraphics[scale=0.33]{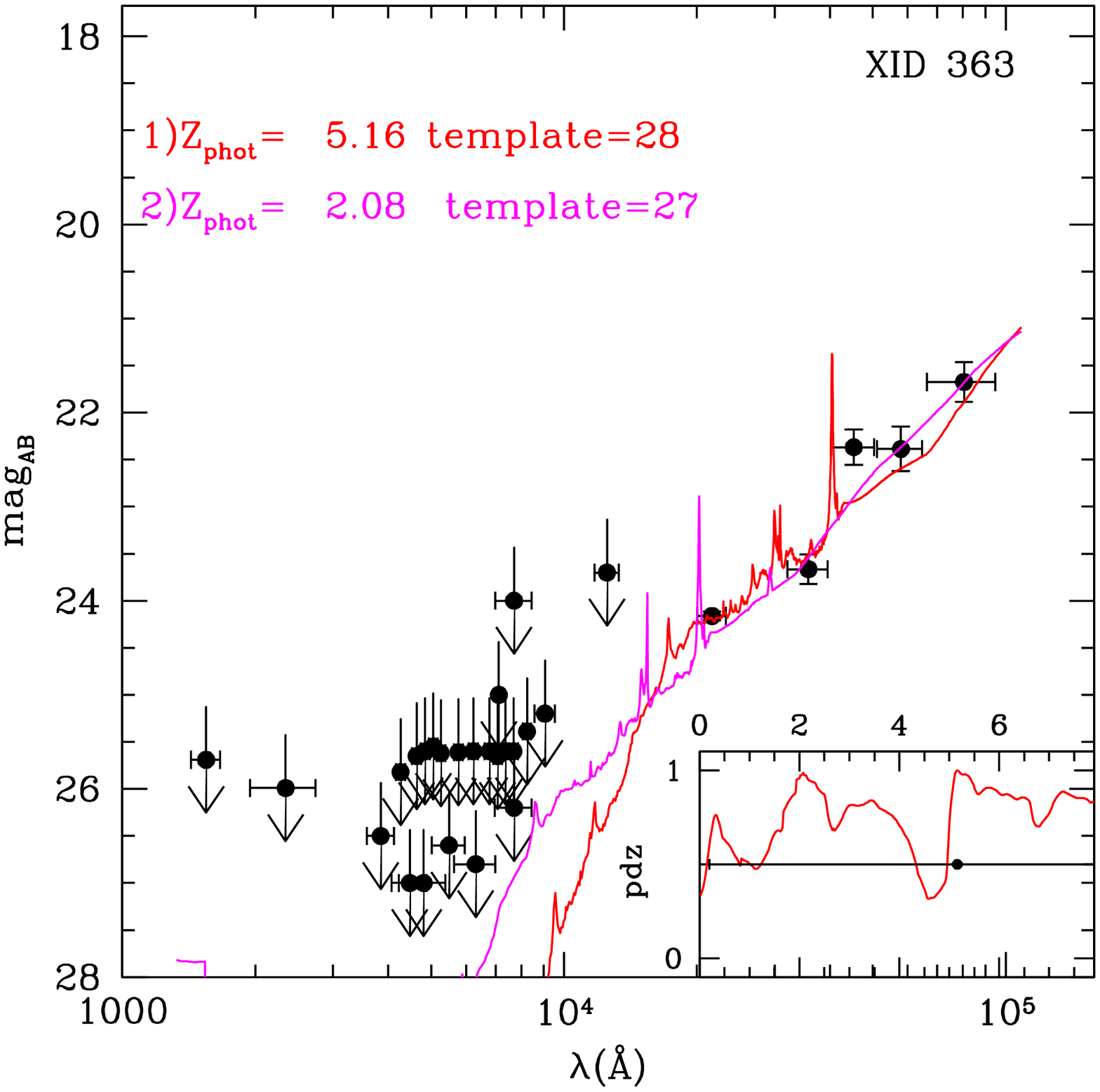} \includegraphics[scale=0.33]{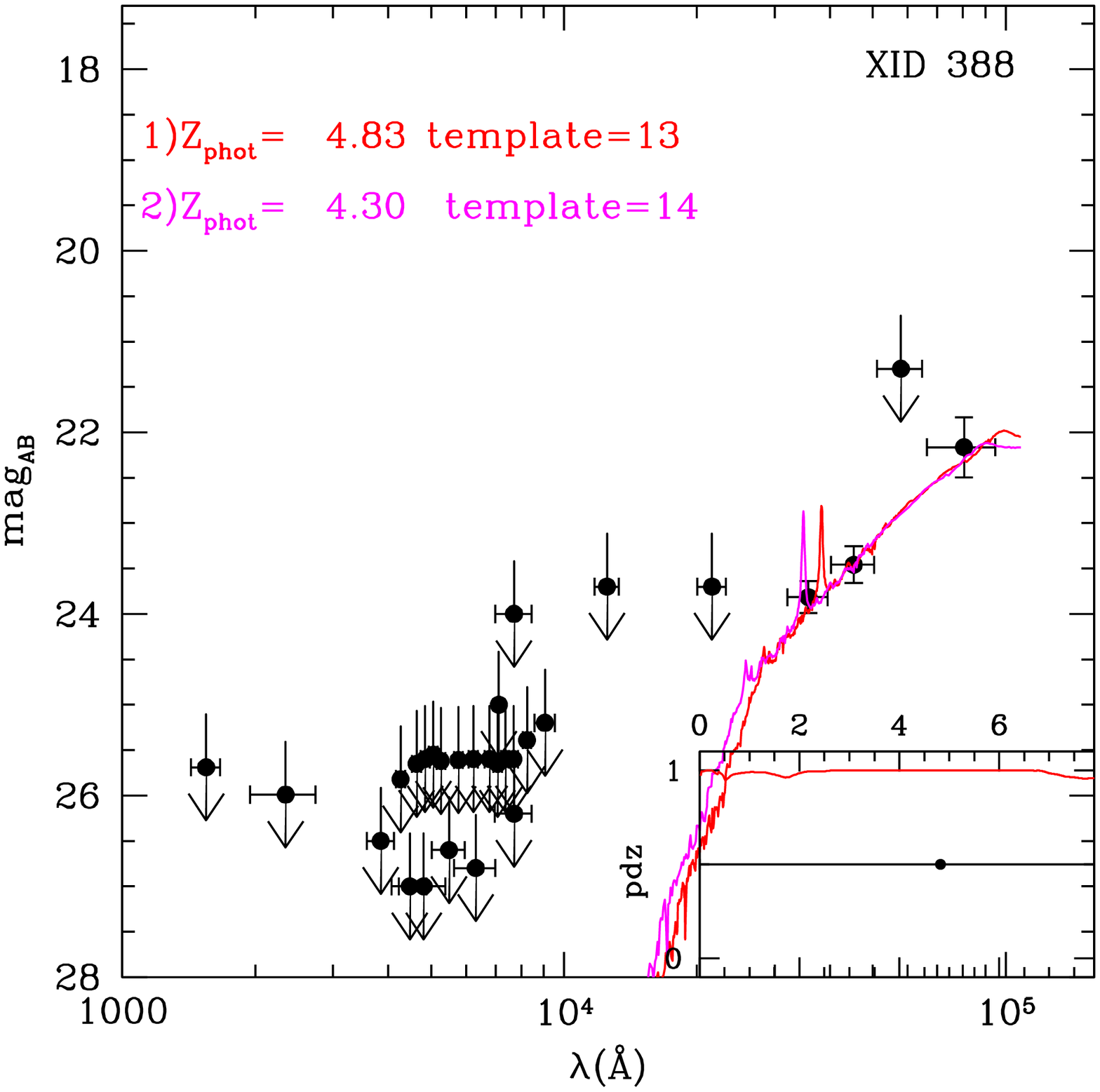}
\includegraphics[scale=0.33]{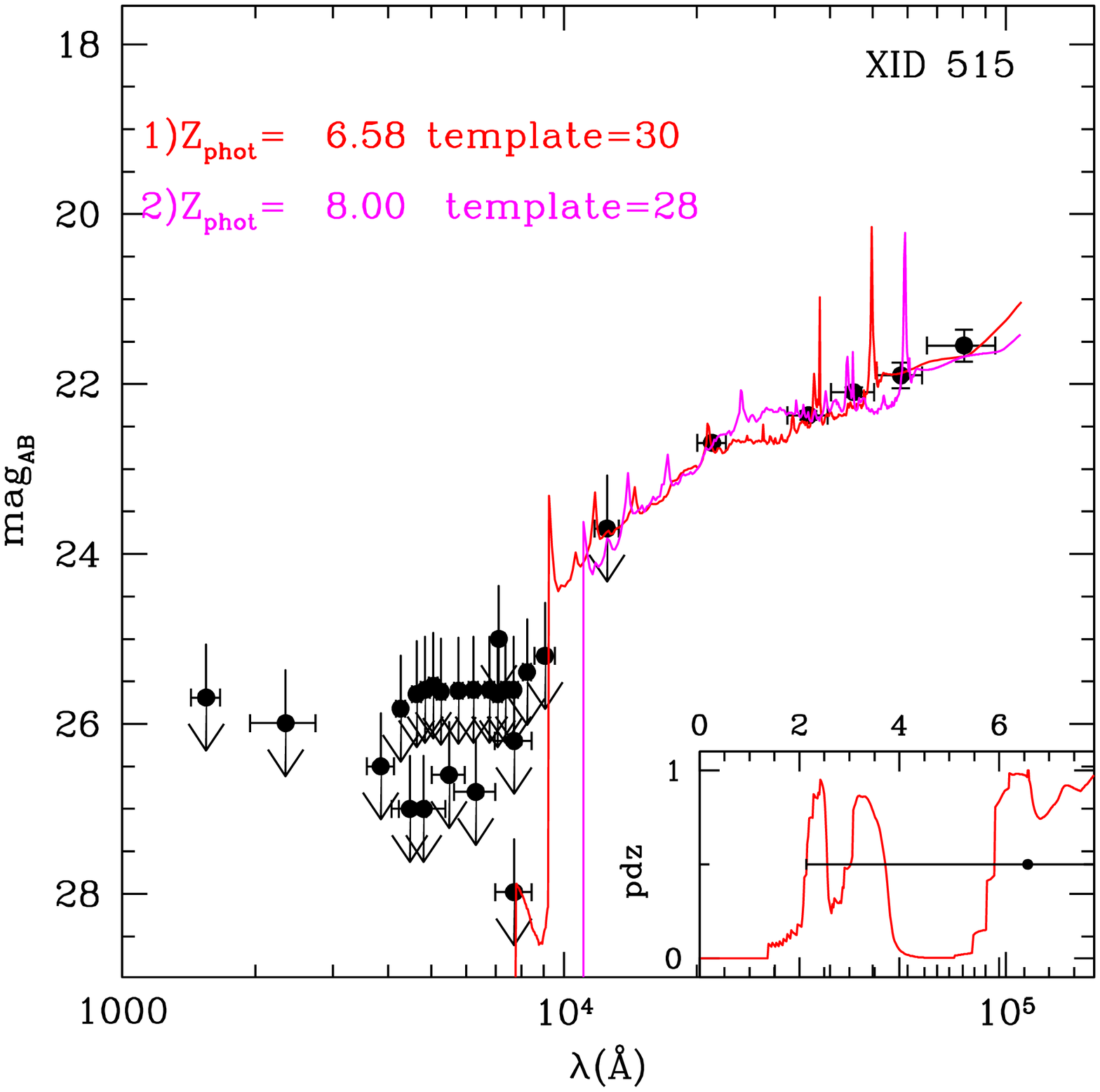} \includegraphics[scale=0.33]{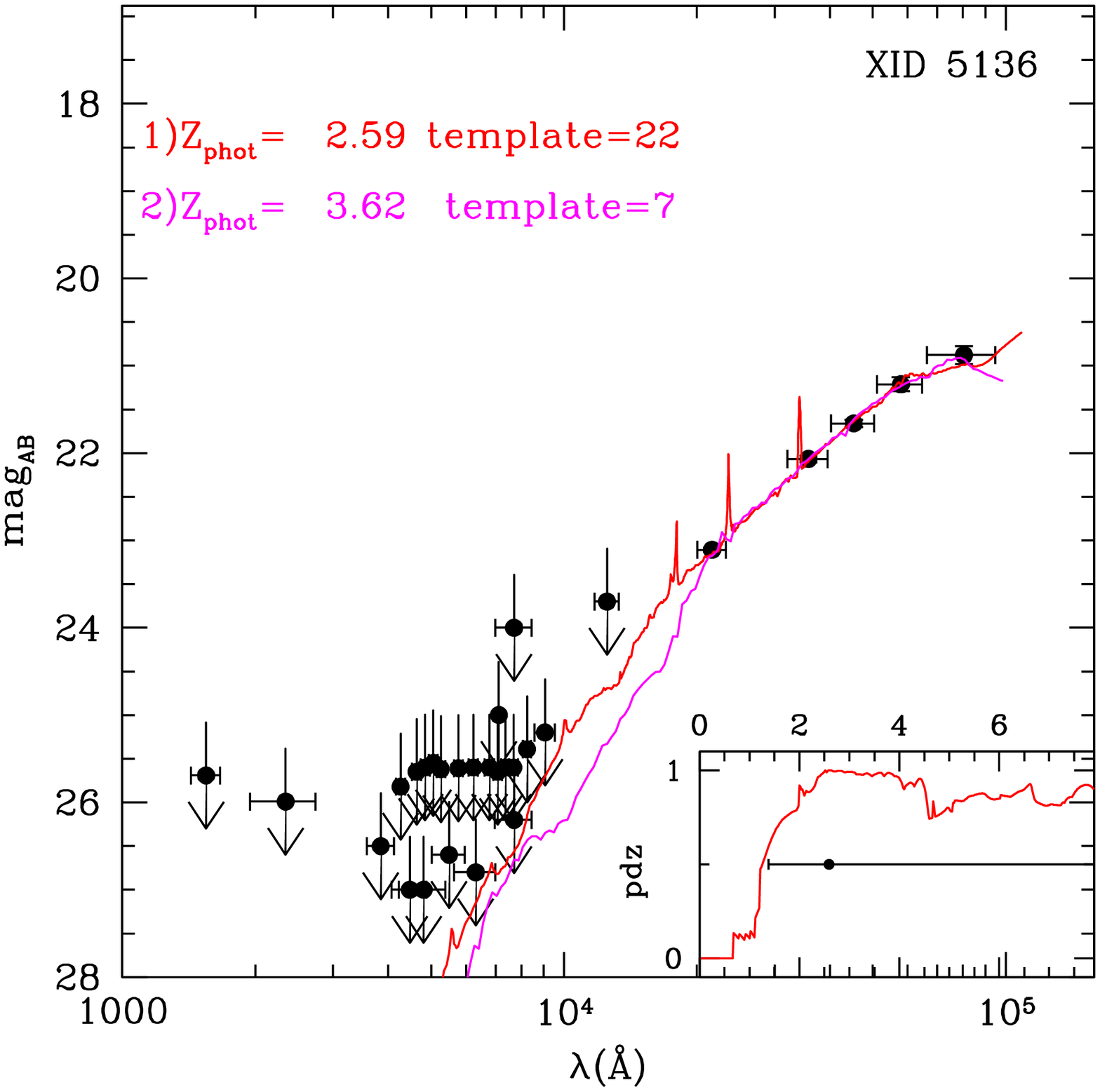}
\includegraphics[scale=0.33]{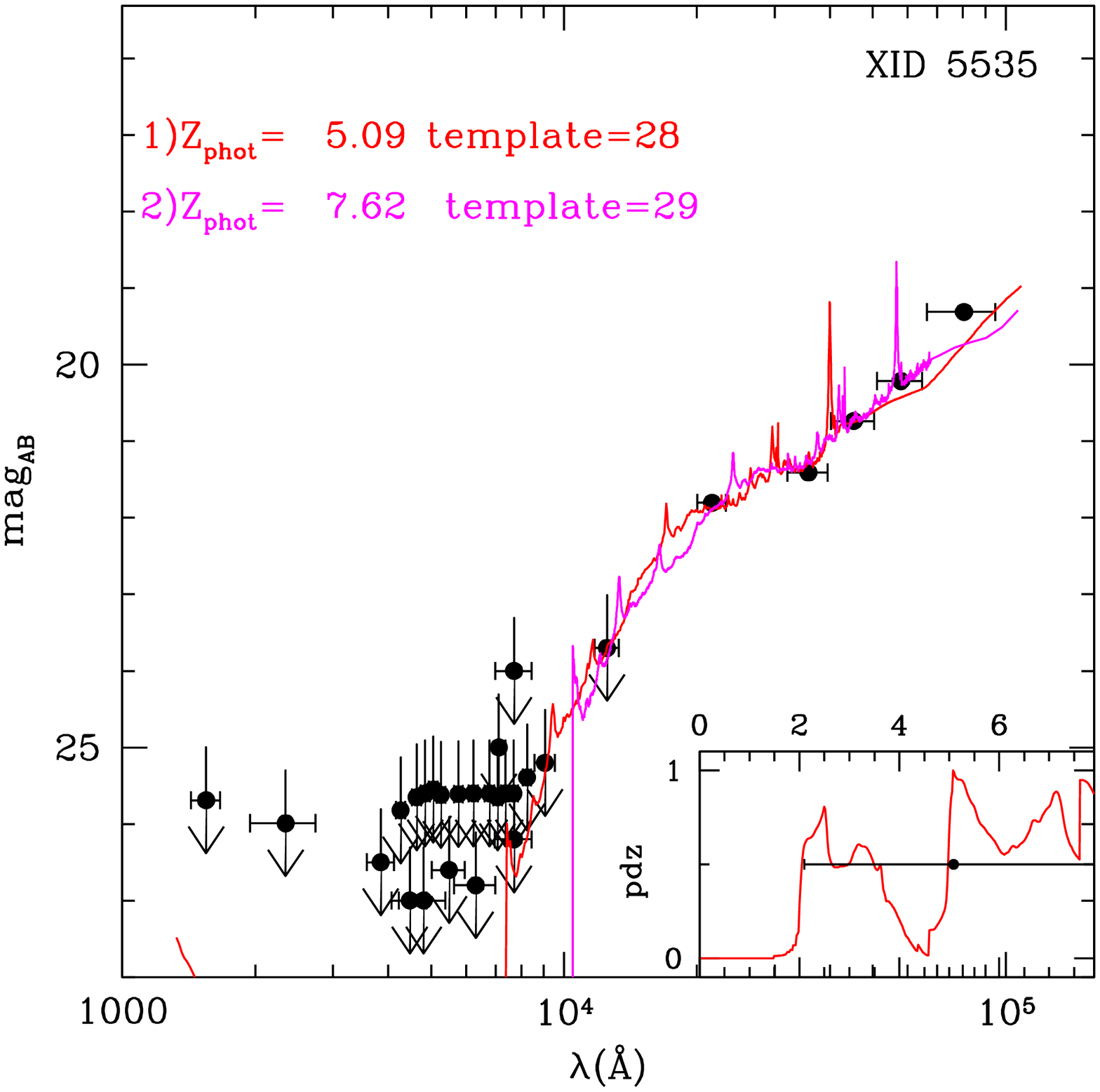} \includegraphics[scale=0.33]{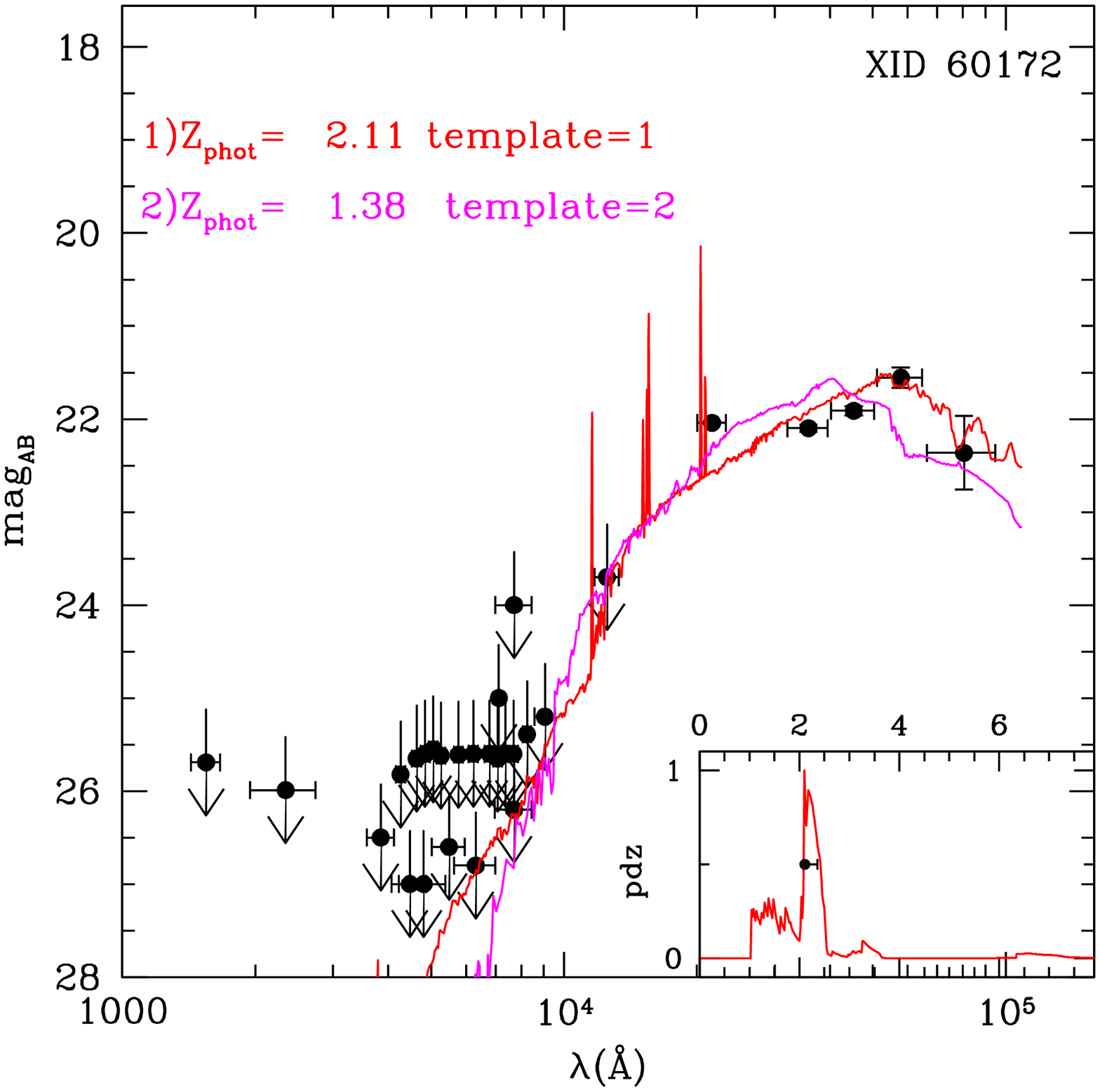}
\caption{{\small SED fitting and photometric redshift estimation for the isolated 6 K band detected sources  for which only 5 (or less)  photometric points are available. The redshift probability distribution is shown in the inserts. Although   the 1-$\sigma$ errors (black horizontal bars in the insert) are so large that a solution at z$\sim$2 can not  be excluded, the best formal fit redshift is greater than 4.}}
\end{center}
\label{fig:faint}
\end{figure*}

Our redshift estimates for  6 of the 10 sources in the {\it
  XMM}-COSMOS optically faint sample is shown in Fig.~\ref{fig:faint}.  We
do not report the results for the remaining 4 sources, as strong
contamination from a nearby object does not allow accurate photometry.
In 4 of the 6 cases, the formal best fit redshift is higher than 4,
but the redshift probability distribution function (see insert in each
figure) indicates that there are not enough constraints to reject a
solution at lower redshift.  In particular, z $\sim$ 2.0-2.5 would be
acceptable for all of them.

\subsection{Reliability of the best fit SED templates}
\label{sec:reliabilitySED}
It is reasonable to ask if the selected SED templates are consistent
with other properties of the sources as revealed in the COSMOS
multi-wavelength data set. For example, 1) are the highly varying
sources mostly fit with a template including a type 1 AGN
contribution?  and 2) were the sources likely to be "host dominated"
on the basis of their low X-ray luminosity better fit by templates
with little overall AGN contribution ?
\begin{figure}[h]
\centering
\includegraphics[scale=0.4]{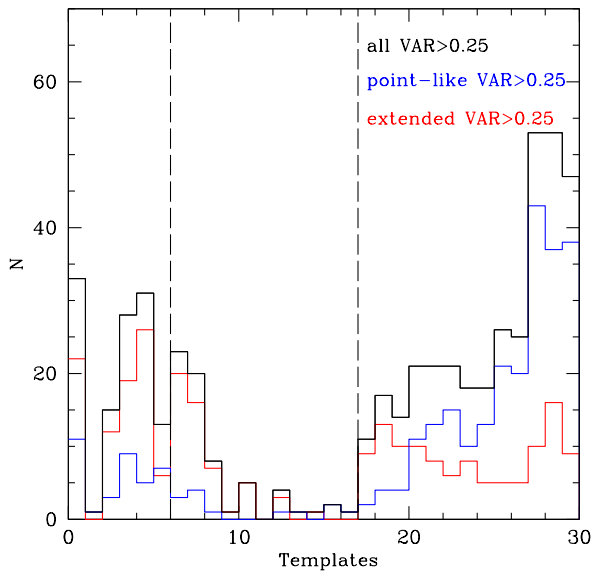}
\caption{Distribution of the best fit SED templates for all
variable sources (black),for the point-like (blue) and the extended (red) sub-samples.
The vertical lines indicate the separation between templates of non-active galaxies (templates 1 to  6), type 2 AGN or with a QSO2 component (templates 7 to 17) and type 1 AGN, QSO or with a QSO component (templates 18 to 30). The bulk of the sources is best fit with a template of the third
group.}
\label{fig:var_mod}
\end{figure}

\begin{figure*}[htdp]
\centering
\includegraphics[scale=0.35]{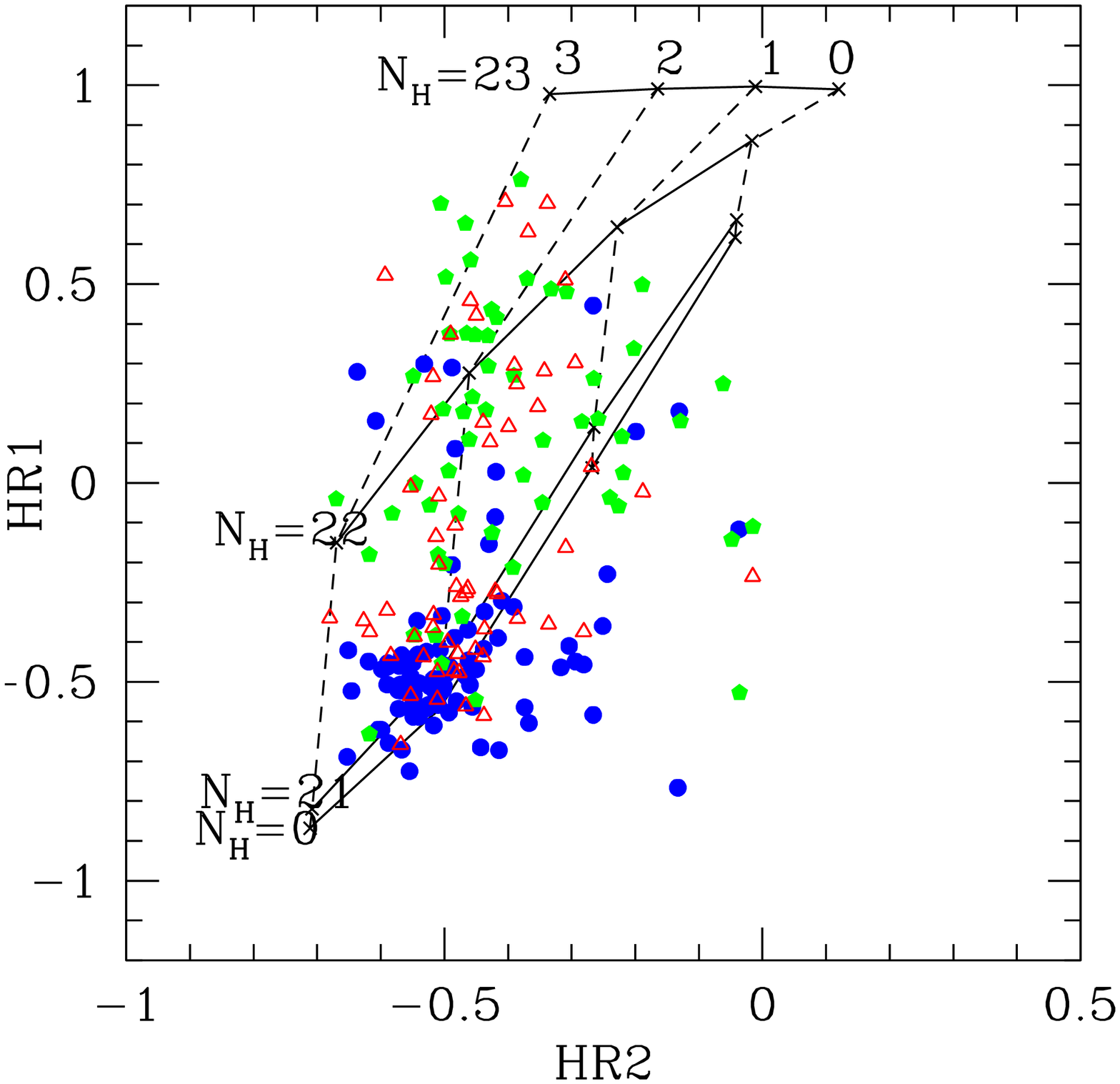}\includegraphics[scale=0.35]{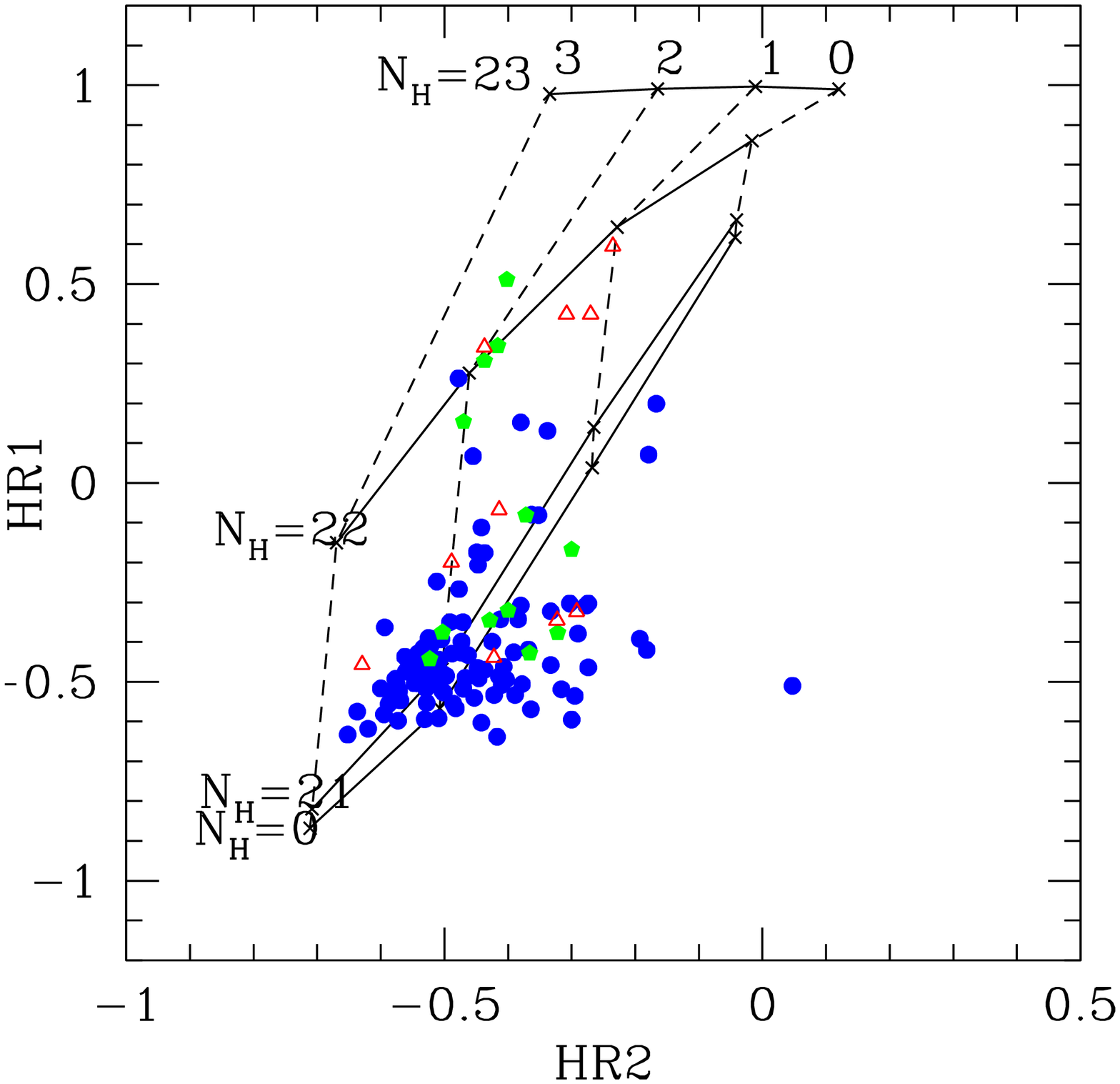}
\caption{Left: X-ray hardness ratios of all XMM-COSMOS non varying sources
($VAR<0.25$) and with  $\Delta$HR1$<0.25$ and $\Delta$HR2$<0.25$. The hardness
ratios are defined as HR1=(F$_{(2-10{\rm keV})}$-F$_{(0.5-2 {\rm
keV})}$)/(F$_{(2-10 {\rm keV})}$+F$_{(0.5-2 {\rm keV})}$) and
HR2=(F$_{(5-10{\rm keV})}$-F$_{(2-10{\rm keV})}$)/(F$_{(5-10{\rm
keV})}$+F$_{(2-10{\rm keV})}$). The best fit templates are color coded
as follows: blue filled circles (Type 1 and QSO1 hybrids -- templates
\#18--30), red open triangles (Type 2 and QSO2 hybrids -- templates
\#7--17) and green open circles (normal galaxies -- templates
\#1--6). The solid lines represent lines of constant N$_{\rm H}$ and
the dotted lines are lines of equal photon index. Right: same as left, for sources with $VAR>0.25$.}
\label{fig:var_mod_Lx}
\end{figure*}

Fig.~\ref{fig:var_mod} clearly shows that most ($\sim$ 62\%) of the
strongly varying sources are best fit with type 1 AGN or QSO dominated
hybrid SEDs while $\sim$ 25\% are best fit with one of the non-active
galaxies.  Similar reassurance is found in the correlation of type 1,
type 2 and normal galaxies with X-ray properties like luminosity and
hardness ratio.  For example, Figure~\ref{fig:var_mod_Lx} shows the
distribution of variable (right panel) and non-variable (left panel)
sources in the $XMM$-color plane. Type 1 AGN are expected to be in a
narrow region with HR1$\sim-0.5$, while type 2 AGN and normal galaxies
have higher HR1 values
\citep{Hasinger:2001zr,Mainieri:2002yq,Della-Ceca:2004mz,Hasinger:2007dn,Cappelluti:2008}.
Our classification is well consistent with the distribution in the
plane of the {\it XMM} sources (see caption of the figure).  Note that
for some sources, the X-ray flux is obtained by stacking up to 9
pointings taken over 3 years.  Thus, the determination of the HR
values can be effected by spectral variability.

\begin{figure}[htbp]
\centering
\includegraphics[scale=0.4]{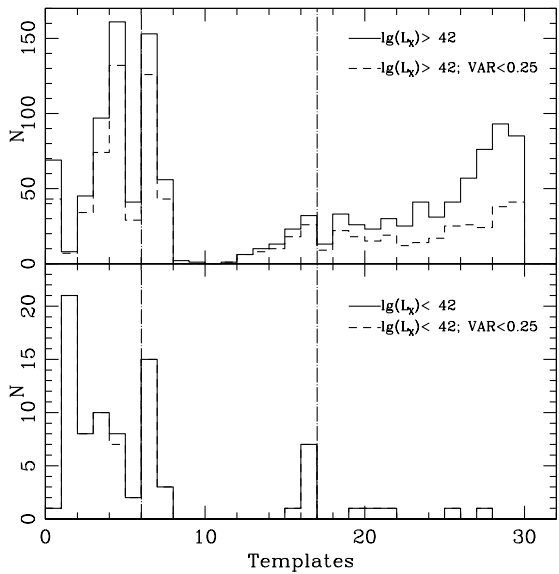}
\caption{SED template distribution for all sources split on
the basis of their X-ray luminosity (bottom: L$_X<10^{42}$\,erg/s, top:
L$_X>10^{42}$\,erg/s) and variability (solid line: all, dashed:
$VAR<0.25$).  Practically all sources with  L$_X<10^{42}$\,erg/s are non-variable (bottom panel).}
\label{fig:lx_var_mod}
\end{figure}

The 2--10\,keV  X-ray luminosity is  commonly used to  distinguish AGN
(L$_{X}>10^{42}$\,erg/s)    from   starburst   or    normal   galaxies
(L$_X<10^{42}$\,erg/s). As we  show in Figure~\ref{fig:lx_var_mod}, the large
majority  of  low-luminosity  sources  are  non-variable  and  a  host
dominated template was selected for  most of them. A large fraction of
high-luminosity  objects  is  instead  varying  and the  best  fit  is
obtained  with  type~1  and  QSO  SEDs.   A  number  of  sources with
(L$_{X}>10^{42}$\,erg/s)  have $VAR<0.25$  and  are fit  by a  normal
galaxy template.  This  is consistent with cases in  which the optical
nuclear activity is completely hidden by the host galaxy.


\subsection{Reliability of the photometric redshift for faint sources}
\label{sec:check_faint}
\begin{figure*}[htbp]
\begin{center}
 \includegraphics[scale=0.35]{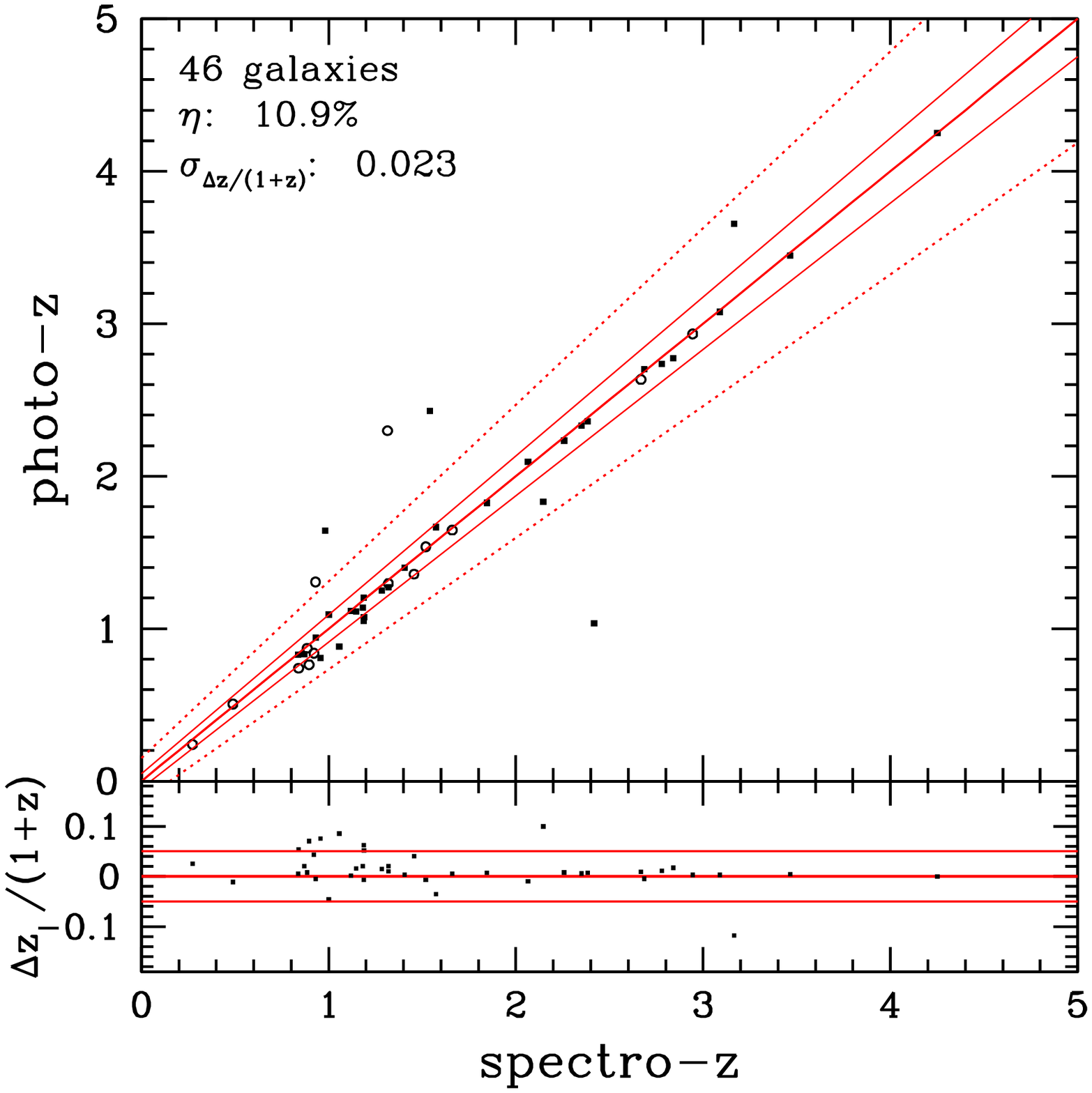} \includegraphics[scale=0.35]{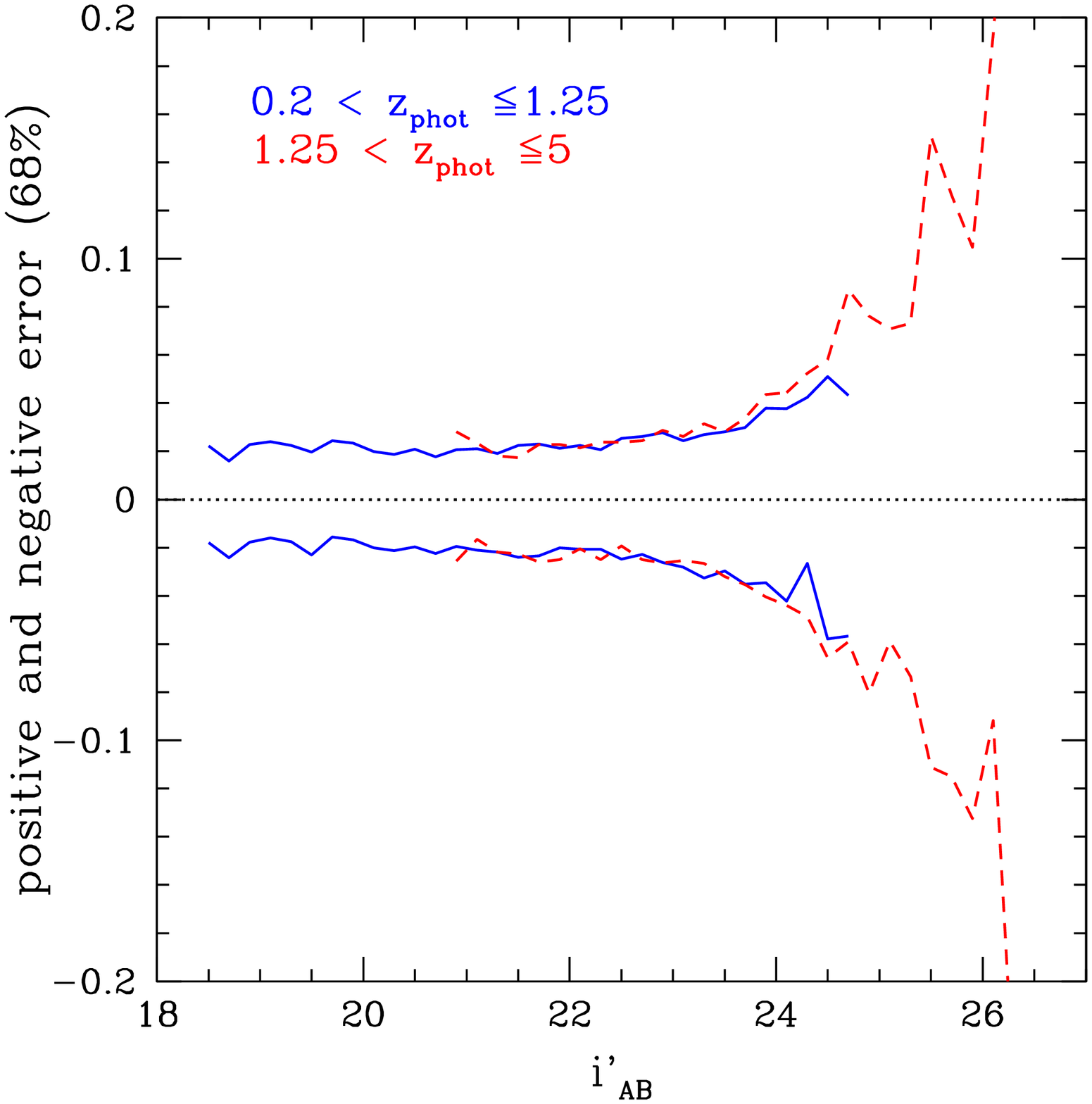} 
\caption{Left:Same as Figure 4  for the 46 sources with i$^+_{AB}>$22.5 with spectroscopic redshifts, which has been used for an {\it a posteriori} test.
Right: 1-$\sigma$ error for the z$_{phot}$ estimate as a function of the apparent magnitude in the redshift
range 0.2 $< z_{phot} \le 1.25 $ (blue) and 1.25 $< z_{phot} \le5 $ (red).}
\end{center}
\label{fig:zp_zs_faint}
\end{figure*}

Recently, the spectroscopic redshift for 46 faint sources
$22.5<$i$_{AB}^*<$24.5 (average value i$_{AB}^*\sim$23.3) became
available.  They are part of a Keck II spectroscopic follow-up of
24$\mu$m selected sources (Kartaltepe et~al.  2008) and part of
$z$COSMOS-{\it faint} (Lilly et~al.  2008).  Fig.~\ref{fig:zp_zs_faint}
(left panel) shows the comparison between the photometric and
spectroscopic redshifts for this sample.  As these sources are faint,
the photometric uncertainties are larger and we should expect lower
photometric redshift accuracy.  In fact, the accuracy become worse
(although still comparable to the typical accuracy of photometric
redshift for non-active galaxies) by a factor of 1.5, from 0.014 to
0.023, while the number of outliers increases from 4.0\% (for the
bright sample) to 10.9\,\%.  However, two of the five outliers (open
circles) do have a secondary maximum in the redshift probability
distribution at the spectroscopic value.  

As the reliability of the photometric redshift based on the comparison
with a spectroscopic sample is limited to the redshift and magnitude
ranges of the the sample, we used the 1$\sigma$ uncertainty (as
derived from the redshift probability distribution function) to test
the accuracy of the photometric redshift for all the {\it XMM}-COSMOS
sources.  Figure~\ref{fig:zp_zs_faint} (right panel) shows the
1$\sigma$ uncertainties in the photometric redshift as a function of
magnitude for sources with 0.2$<$z$<$1.25 and 1.25$<$z$<5$.  Unlike
normal galaxies (see, for example, Figure 9 in Ilbert et~al. 2008),
the uncertainties for the two redshift ranges behave in the same way
and up to i$^+_{AB}\sim$24 remain almost constant, of the order of
0.02. Only beyond i$^+_{AB}>$24 the uncertainties increase and reach
an amplitude of 0.2 at i$^+_{AB}\sim$26.  This is expected as AGN show
prominent emission lines, all over the spectrum. In particular, beside
the well known optical lines (H$_{beta}$, [OIII], H$_{alpha}$, etc.)
two strong emission lines in the blue part of the spectrum (CIV[1549\AA] and
MgII[2798\AA]) are available at high redshift.  The COSMOS survey, dense with
intermediate band photometry, allows us to identify these lines and
thus provides a reliable photometric redshift estimate over the full
redshift range. The reliability of our photometric redshift is
diminished only by the larger photometric errors at faint magnitudes.

\subsection{Reliability of star/galaxy separation}
\label{sec:star_gal}
In addition to galaxy and AGN templates, stellar templates have been
used to identify stars in the $XMM$-COSMOS sample. We constructed a
spectral library of low mass stars \citep{Chabrier:2000zl} and
sub-dwarf O and B stars, white dwarfs and binary systems
\citep{Bixler:1991yg}. The reliability of the star/galaxy separation
was tested on the sub-sample with spectroscopic identification.  We
defined as star each point-like source with
\begin{equation}
\chi^2_{star}\leq\chi^2_{qso}
\end{equation}
and the remaining as extra-galactic.  In this way we classify  46
sources as stars, 28  of which are already spectroscopically confirmed.
In summary, 439 (out of 396+46=442) galaxies and 24 (out of
28) stars were classified  correctly.  The completeness in identifying
galaxies  is 439/442= 0.99  and   24/28= 0.86 for  the stars.
Two of  the three misclassified galaxies were  missing IRAC photometry
and thus the possibility to  disentangle between star and galaxies was
limited.
  
\section{Release of the photometric redshift catalog}
\label{sec:catalog}
An  electronic  version  of  the  complete  photometric  redshift  and
classification catalog  of the  {\it XMM}-COSMOS sample  is available.
For  each source  the coordinates,  the photometric  redshift  and the
classification  in  star/galaxy  is  provided.  A  more  comprehensive
catalog is available by request\footnote{ms@astro.caltech.edu}.
   
\section{Discussion}
\label{sec:discussion}
 In this section we analyze the impact of each contribution (new
  template library, correction for variability, photometric coverage)
  on our final result.  We show that while the number of outliers can
  be decreased with the appropriate SED template library and the
  correction for variability, a good photometric coverage is essential
  in order to improve the accuracy of the photometric redshifts.

\subsection{Relevance of SED templates library}
\label{sec:dislib}
 An X-ray selected sample of sources consists either of low redshift
 star-forming galaxies or AGN and QSOs and the choice of templates
 must represent these populations.  To verify this, we first compared
 our photometric redshifts with those obtained for the same sources by
 Ilbert et al. (2008) using the same photometric data set but only
 normal galaxy SED templates.  The result is shown in the first row of
 Table ~\ref{tab:SEDvar}.  For the ``etxnv'' sample, where the sources
 are galaxy dominated, the library of Ilbert et al.  (2008) provides
 an excellent accuracy although the number of outliers is large.  For
 the ``qsov'' sample dominated by AGN and QSOs, the lack of
 appropriate templates leads to an expected failure\footnote{In Ilbert
   et~al. (2008) the new publicly available COSMOS photometric
   redshift catalog is presented. This catalog supersedes all previous
   ones (Mobasher et~al. 2007, Ilbert et al. 2006) and is composed of
   600,000 sources with i$^+_{AB}<$26 and not detected in
   X-rays. In order to decrease the risk of degeneracies for this
   sample of normal galaxies, no AGN/QSOs templates were used to
   compute the photometric redshifts.}.

The addition of AGN and QSO templates in the library does not
automatically imply an improvement in the photometric redshifts.  We
verified this by computing the photometric redshift using the library
of Polletta et al. (2007) which includes normal galaxies, AGN and
QSOs.  The result of the test is shown in the second row of Table
~\ref{tab:SEDvar}.  In this case, the accuracy is very good and
comparable with the one obtained with our library. However, the number
of outliers remain high.  We then conclude that our library,
constructed as described in $\S $~\ref{sec:templates}, can be considered as
representative of a typical population of X-ray emitting galaxies.

\subsection{Relevance of correction for variability}
\label{sec:disvar}

In the previous section we have shown the importance of the library to
increasing the accuracy and to reducing the number of outliers. The
comparison has been performed on the photometric data set corrected
for variability; thus, it is natural to ask how much the improved
performances shown by our library are depending on it.  In the third
row of Table ~\ref{tab:SEDvar}, our library has been used on the
photometric data set without any correction for variability.  While the
accuracy of the photometric redshift did not change, the number of
outliers is increased by a factor of 2 in both ``extnv'' and ``qsov''
sub-samples, demonstrating the importance of the correction for
variability.
\begin{table*}[htdp]
\caption{Relevance of variability correction and template library}
 \begin{center}
\begin{tabular}{c|c|c|c|c}
 &\multicolumn{2}{c}{extnv} & \multicolumn{2}{c}{qsov}\\
 & $\sigma_{\Delta_{z}/(1+z)}$ &    outliers (\%) &  $\sigma_{\Delta_{z}/(1+z)}$ &    outliers (\%)\\
\hline
\hline
                                                 &        &         &            &       \\
Ilbert08\tablenotemark{1}  &0.014&14.0& 0.419 &71.4\\ 
                                                 &        &         &            &       \\
\hline
                                                 &        &         &            &       \\
Polletta07\tablenotemark{2} + var. corr.   &0.020 & 12.8 & 0.014 & 15.0\\
                                                 &        &         &            &       \\
\hline
                                                 &        &         &            &       \\
Salvato08\tablenotemark{3}, no var. corr. &0.017 & 4.1 &0.013 & 11.6 \\
                                                 &        &         &            &       \\
 \hline
                                                 &        &         &            &       \\
Salvato08\tablenotemark{3} +  var. corr.   &0.019 & 2.3  & 0.012 & 6.3\\
                                                 &        &         &            &       \\
\end{tabular}
\end{center}
\tablenotetext{1}{{Library} used in Ilbert et~al. 2008}
\tablenotetext{2}{{Library} used in Polletta et~al. 2008}
\tablenotetext{3}{{Library} used in this work}
\label{tab:SEDvar}
\end{table*}

 \subsection{Relevance of spectral coverage}
\label{sec:discov}
To quantify the benefits of  spectral coverage, we split the available
photometric  data into 5  sub-sets ({\it  GALEX}, optical  broad band,
optical intermediate band, near-infrared,  and IRAC). Next, we derived
the  photometric redshifts for  the "extnv"  and "qsov"  samples using
different combinations  of these sub-sets and compared  the results in
terms of  accuracy and number  of outliers. We  find that for  all the
combinations  the  accuracy  is $\sigma_{\Delta  z/(1+z_{spec})}<0.07$,
which  is good  enough for  most AGN  studies. A  summary is  given in
Table~\ref{tab:summary}.

For the "extnv" sample, the by  far best result was achieved using the
entire  spectral coverage.   This is  not surprising  and  is also 
discussed for  normal galaxies in COSMOS  (Ilbert et~al. 2008).  The 30
band  photometry  of  COSMOS  essentially resembles  a  low-resolution
spectrum and  a representative  SED template library  can be  fit with
high reliability. 

Table~\ref{tab:summary} shows also  the importance of the intermediate
optical  filters  which allow  the  precise  localization of  spectral
features  (e.g., emission  and  absorption lines,  Balmer break).  The
elimination of  these filters produces a  significantly worse accuracy
(by  a factor  of  3  for both the  "qsov" and the   "extnv" samples).
 At the same time the number of outliers increases by a factor
of $\sim$ 3 for both samples.

\begin{table*}[htpb]
\caption{Relevance of spectral coverage and photometric resolution  for the variability corrected bright spectroscopic sample. }
\begin{center}
\begin{tabular}{c|c|c|c|c|c|c|c}
Sample & GALEX & Optical& Optical &NIR&IRAC&\multicolumn{2}{c}{Reliability}\\
                &               &Broad Band  &                      IB \& NB                         &      &         &   outliers (\%) & $\sigma_{\Delta_{z}/(1+z)}$         \\
\hline
      & & & & & & &\\
qsov  & & & & & & 6.3 & 0.012\\
      & $\surd$ & $\surd$ & $\surd$ & $\surd$ & $\surd$ & &  \\
extnv & & & & & & 2.3 & 0.019\\
      & & & & & & &  \\
\hline
      & & & & & & &\\
qsov  & & & & & & 19.4 & 0.063\\
      & $\surd$ & $\surd$ & -- & $\surd$ & $\surd$ & &  \\
extnv & & & & & & 6.8 & 0.056\\
      & & & & & & &  \\
\hline
      & & & & & & &\\
qsov  & & & & & & 6.9 & 0.011\\
      & $\surd$ & $\surd$ & $\surd$ & $\surd$ & -- & &  \\
extnv & & & & & & 5.9 & 0.019\\
      & & & & & & &  \\
\hline
      & & & & & & &\\
qsov  & & & & & & 14.3 & 0.012\\
      & $\surd$ & -- & $\surd$ & $\surd$ & $\surd$ & &  \\
extnv & & & & & & 6.8 & 0.021\\
      & & & & & & &  
\end{tabular}
\end{center}
\label{tab:summary}
\end{table*}


In contrast, the removal of the broad optical bands photometry when
the intermediate bands are available, does not change significantly the
accuracy of the photometric redshift, neither for the "qsov" nor for
the "extnv" samples. However, in both cases the number of outliers
increases dramatically.  The same is happening when the IRAC photometry
is not considered for the "extnv" sample.  As already pointed out by
\cite{Rowan-Robinson:2008ph} the presence of IRAC photometry is
crucial in suppressing the number of outliers.

\section{Conclusions}
\label{sec:conclusions}
 In   this   paper  we   presented   the   photometric  redshifts   and
classifications  for  1542   {\it  XMM}-COSMOS  sources  with  optical
counterpart.   Using SED  matching to  up to  30 photometric  bands we
identified 46 Galactic  foreground stars, 464 sources best  fit with a
non-active galaxies template and 1032 best fit by a template which has
an  AGN contribution.  A high  reliability in  recognizing  stars from
galaxies and AGN was demonstrated using a spectroscopically identified
sub-set of  470  sources.  Overall, we  reach a photometric
redshift accuracy of   $\sigma_{\Delta z/(1+z_{\rm spec})}=0.014$
(i$^+_{\rm AB}<22.5$)  with  4.0\,\% outliers.  However, the most
important result  is perhaps  the accuracy for  the sub-sample  of AGN
dominated  sources ($\sigma_{\Delta z/(1+z_{\rm  spec})}=0.011$) which
marks  an significant improvement  in photometric redshift estimation for this kind of sources.
 The number of  outliers, however, is still non-negligible (6.3\% when considering  the "qsov" sample alone).

The excellent photometric redshifts and classifications have been made
possible by a number of factors.  Here, the most important was likely
the large spectral coverage from UV to 8\,$\mu$m, densely sampled with
broad, intermediate and narrow filters.  The 25 optical bands
essentially mimic low resolution spectra and are powerful in
constraining the positions of spectral features.  The
near/mid-infrared coverage provides the spectral range required to
better distinguish stars from galaxies and helps to constrain SED
templates, especially for passive galaxies.

Also important was the recognition of and correction for
variability. This is especially relevant for AGN dominated sources,
for which photometry collected over a large time span may dramatically
vary from a snapshot SED.  We applied a correction for variability by
re-scaling the optical photometry to a common epoch close to that of
IRAC observations.

The correction decreased the number of outliers by a factor of 2.  The
significant improvement obtained in our sample with this correction
demonstrates the importance of quasi-simultaneous observations for
projects aiming for good photometric redshift estimates of AGN. Future
surveys will have to consider this point very carefully.  Deep imaging
in a common filter in each observing epoch would provide the
possibility of a valuable first order correction.

Finally, we used a revised set of SED templates to fully represent the
entire range of different contributions of active galactic nuclei and
their host galaxies. Empirical SED templates are often obtained from
objects in the nearby Universe where the nuclear and extended fluxes
are easily resolved.  Their application to high-redshift objects,
where the two contributions are mixed together, may not always be
appropriate.  Our hybrid templates already showed a significant
improvement, although further advancements are required.

The catalog of {\it XMM}-COSMOS selected sources with photometric
redshifts and classifications, together with the X-ray properties
(Brusa et~al.  2008) offers an unprecedented possibility to study the
luminosity function and redshift distribution of AGN. In combination
with the immense multi-band COSMOS data set and the new photometric
redshifts for optically selected sources (Ilbert et~al. 2008), studies
of the environment of AGN will also allow to probe AGN fuelling
models, clustering and evolution.

\acknowledgments 
We gratefully acknowledge the contributions of the
entire COSMOS collaboration consisting of more than 100 scientists.
More information on the COSMOS survey is available at
http://www.astro.caltech.edu/cosmos. We acknowledge the anonymous
referee for helpful comments that improved the paper. We also
acknowledge the hospitality of the Institute for Astronomy, Hawaii,
and the support of the Aspen Center for Physics where the work started
and the manuscript was completed.  GH acknowledges support by the
German Deutsche Forschungsgemeinschaft, DFG Leibniz Prize (FKZ HA
1850/28-1).  This work was supported in part by NASA Grant GO7-8136A.
\bibliographystyle{apj}
\bibliography{../../newlibrary}

\end{document}